\documentclass[12pt]{article}
\usepackage{graphicx}
\usepackage{latexsym}
\usepackage{epsfig}
\usepackage{cite}
\usepackage{amssymb,amsmath}
\usepackage{enumitem}

\title{Hadronic Parity Violation}
\author{Wick C. Haxton\\ 
Department of Physics, MC7300\\
University of California, Berkeley, and\\
Lawrence Berkeley National Laboratory\\
Berkeley, CA  94720\\
and\\
Barry R. Holstein\\
Department of Physics-LGRT\\
University of Massachusetts\\
Amherst, MA  01003}
\begin{titlepage}
\begin{document}
\maketitle
\begin{abstract}
The history and phenomenology of hadronic parity nonconservation (PNC) is
reviewed.  We discuss the current status of the experimental tests and theory.
We describe a re-analysis of the asymmetry for $\vec{p}+p$ that, when combined with
other experimental constraints and with a recent
lattice QCD calculation of the weak pion-nucleon coupling $h_\pi^1$, reveals a
much more consistent pattern of PNC couplings.
In particular, isoscalar coupling strengths are similar to but somewhat larger
than  the ``best value" estimate of Donoghue, Desplanques, and Holstein,
while both lattice QCD and experiment indicate a suppressed $h_\pi^1.$
We discuss the relationship between meson-exchange
models of hadronic PNC and formulations based on effective theory, stressing
their general compatibility as well as the challenge presented to theory
by experiment, as several of the most precise measurements involve 
significant momentum scales.  Future directions are proposed.
\end{abstract}
\end{titlepage}
\section{Introduction}
The experimental study of parity nonconservation (PNC) in $\Delta S=0$ hadronic interactions began only a year after Lee and Yang pointed out that parity might be violated in the weak interaction\cite{ley}.  In 1957, the year that the ${}^{60}$Co beta decay experiment of Wu et al. confirmed the Lee and Yang hypothesis\cite{amb}, Neil Tanner published the (negative) results of an experiment seeking to measure PNC in the reaction ${}^{19}$F(p,$\alpha){}^{16}$O \cite{nta}.  Although the Tanner  experiment lacked the sensitivity required to observe a signal, his effort  was followed by others, with experiments designed to isolate tiny weak effects within
systems where strong and electromagnetic interactions dominate continuing until today.  The presence of weak effects is apparent from results such as the $2\%$ photon asymmetry in the electromagnetic decay of an isomer of ${}^{180}$Hf\cite{hfe}
 \begin{equation}
A_\gamma({}^{180}{\rm Hf}^*\rightarrow{}^{180}{\rm Hf}+\gamma)=-(1.66\pm 0.18)\times 10^{-2}\label{eq:xz}
\end{equation}
or the nearly $10\%$ asymmetry in the scattering of longitudinally polarized neutrons from $^{139}$La\cite{lae}
\begin{equation}
A_h(\vec{{\rm n}}+{}^{139}{\rm La})=(9.55\pm 0.35)\times 10^{-2}\label{eq:xw}.
\end{equation}
The asymmetries given in Eqs. (\ref{eq:xz}) and (\ref{eq:xw}) are, however, anomalously large, amplified by nearly degenerate nuclear states having the same spin but opposite parity.  Indeed the natural scale of hadronic PV effects is five orders of magnitude smaller, $\sim G_F F_\pi^2 \sim 10^{-7}$, where $F_\pi \sim 92.4$ MeV is the pion decay constant.  In the more than half century since the discovery of PNC, many searches for hadronic PNC effects have been performed, both with and without the use of nuclear amplification, creating a substantial body of results. (See, for example, the reviews of \cite{rev1,rev2,rev3}).  

The focus of such work has changed from detection to analyzing the structure of the PNC response.   At the time of the first experiments, only the charged weak current (CC) was known to exist,
\begin{equation}
J_\mu^+=\cos\theta_C\bar{u}\gamma_\mu(1+\gamma_5)d+\sin\theta_C\bar{u}\gamma_\mu(1+\gamma_5)s
\label{eq:CC}
\end{equation}
where $\theta_c\sim 13^o$ is the Cabibbo angle and $\gamma_5=-i\gamma^0\gamma^1\gamma^2\gamma^3$ is the negative of that defined by Bjorken and Drell \cite{bd}.  The $\Delta S=0$ weak interaction resulting from the contraction of this current and its Hermitian conjugate,
\begin{equation}
{\cal H}_w^{CC}={G_F\over \sqrt{2}}J_\mu^{+\dagger}J^{+\mu}+{\rm h.c.}
\end{equation}
then consists of two components, a $\Delta I=0,2$ piece multiplied by $\cos^2\theta_C\sim 1$ which results from the symmetric product of isotopic spin $I=1$ parts of the weak current and a $\Delta I=1$
 piece multiplied by $\sin^2\theta_C\sim 0.04$ which results from the symmetric product of $I=1/2$ parts of the current.  Thus the $\Delta I=1$ piece of the strangeness conserving PNC weak interaction generated by the current of Eq. ({\ref{eq:CC}) is strongly suppressed with respect to its $\Delta I=0,2$ counterparts.  Early on it was recognized that experimental studies of this $\Delta I =1$ component could be a probe possible neutral weak currents (NC).  The  development of the standard model and the experimental discovery of the hadronic NC
\begin{equation}
J_\mu^{0}=\bar{u}\gamma_\mu(1+\gamma_5)u-\bar{d}\gamma_\mu(1+\gamma_5)d-4\sin^2\theta_\mathrm{W}J_\mu^{em},
\end{equation}
where $\theta_\mathrm{W}$ is the Weinberg angle and $J_\mu^{em}$ the electromagnetic current,
led to an important program of semi-leptonic weak interaction studies of this current.  In the case of hadronic
weak interactions, this current generates a Hamiltonian
\begin{equation}
{\cal H}_w^{NC}={G_F\over 2\sqrt{2}}J_\mu^{0\dagger}J^{0\mu}+{\rm h.c.}.
\label{eq:NC}
\end{equation}
that does add an unsuppressed NC contribution to the $\Delta S=1$ interaction: the product in Eq. (\ref{eq:NC}) generates $\Delta I=0,1,2$ interactions.  
As NCs do not contribute to flavor-changing interactions, it was recognized that the resulting opportunity to study the NC hadronic weak interaction in $\Delta S=0$ systems -- particularly $\Delta I=1$ channels --  is unique. The nucleon-nucleon and nuclear systems are the only practical arenas for such studies, with PNC then available as a tool for isolating weak effects in systems dominated by strong and electromagnetic interactions.  Summaries of experimental work are given in Sec. 2\cite{rev1,rev2,rev3}.

During the past five decades a great deal of theoretical work has also been performed, with the goal of understanding the patterns of observation
and non-observation of PNC that emerged from the experiments.  We describe several of the important unresolved issues in our description of PNC in Sec. 3, using the traditional meson-exchange picture of hadronic weak interactions.  In Sec. 4 discuss issues in the context of partial-wave analyses and effective field theory, 
approaches that allow in principle for a more systematic approach to PNC (though their description of the weak 
nucleon-nucleon (NN) 
interaction is largely equivalent to the meson-exchange formulation).  Then in Sec. 5, we describe an envisioned program that would have the potential to clear up many of the existing uncertainties.  Our paper concludes with a brief summary in Sec. 6.

\section{Experimental Summary}
In the more than five decades since publication of the Tanner paper, many additional experiments have been carried out to probe the  hadronic weak interaction.  
Ideally one would do a series of measurements in the NN system, where the strong interaction physics can be handled
precisely: in the Danilov amplitude decomposition we describe in Sec. 4, a minimum of five measurements would be 
needed to determine the full set of S-P amplitudes.  As the natural scale of NN PNC observables is $\sim 10^{-7}$, exceptional effort must be invested to accumulate sufficient counts to see an effect and to control systematics that could yield false signals. 
The results to date include 
\begin{itemize}
\item[a)] The analyzing power for the scattering of longitudinally polarized protons from an unpolarized proton target has been measured at 13.6\cite{ghs} and 15\cite{nagle} MeV by groups from Bonn and Los Alamos, and at 45 MeV\cite{ght} by a group from PSI.  Furthermore,  a medium energy measurement of $A_L$ was made at TRIUMF\cite{TRIUMF}.  The results are
    \begin{eqnarray}
    A_L(\vec{{\rm p}}{\rm p}; 13.6\,\,{\rm MeV})&=&(-0.93 \pm 0.20\pm 0.05)\times 10^{-7} \nonumber \\
    A_L(\vec{{\rm p}}{\rm p}; 15\,\,{\rm MeV})&=&(-1.7 \pm 0.8)\times 10^{-7} \nonumber \\
        A_L(\vec{{\rm p}}{\rm p}; 45\,\,{\rm MeV})&=&(-1.57 \pm 0.23)\times 10^{-7} \nonumber \\
        A_L(\vec{{\rm p}}{\rm p}; 221\,\,{\rm MeV}&=&(+0.84 \pm 0.34)\times 10^{-7}
    \end{eqnarray}
\item[b)] The asymmetry in the radiative capture of cold polarized neutrons on a parahydrogen target, $np\rightarrow d\gamma$, was measured at Grenoble in 1977 \cite{cavaignac} and much more recently at LANL \cite{bow}.  The two results are of comparable sensitivity, with both failing to find a non-zero signal for PNC,
\begin{eqnarray}
A_\gamma(\vec{{\rm n}}{\rm p})|_\mathrm{Grenoble}&=&(0.6\pm 2.1)\times 10^{-7} \nonumber \\
A_\gamma(\vec{{\rm n}}{\rm p})|_\mathrm{LANL}&=&(-1.2\pm 1.9\pm 0.2)\times 10^{-7}.
\end{eqnarray}

\item[c)] The circular polarization of the 2.22 MeV photon emitted in the capture of unpolarized thermal neutrons by protons was
measured at the Leningrad reactor, yielding the upper bound \cite{lob1,lob2}
\begin{equation}
P_\gamma({\rm np})=(1.8\pm 1.8)\times 10^{-7}
\end{equation}
\end{itemize}

Because the only definitive measurement of PNC in the NN system comes from the $\vec{\mathrm{p}}$p system, experimentalists
have turned to few-body systems.  Here calculations can be done in principle with quasi-exact 
nonperturbative strong interaction methods, though
applications to scattering states are far less developed than to bound states.  Faddeev-Yakubovsky methods and variational methods
in hyperspherical harmonic bases can and have been applied to A=4 systems: the first bench-mark PNC calculations for ${}^3$He(n,p)$^3$H
have been recently reported \cite{viviani}.  Hyperspherical harmonic and quantum Monte Carlo methods are under active development for A=5 PNC applications
such as the spin rotation of polarized neutrons in helium \cite{viviani,nollett}.  Thus within the next few years, quasi-exact methods should exist for extracting
weak couplings from few-body systems.  In the interim, model-based scattering calculations can be done with bound-state wave functions that are
effectively exact.  Few-body experiments include:
\begin{itemize}
\item[d)] A PSI experiment measured the cross section difference in the scattering of longitudinally polarized protons from a
${}^4$He target at 46 MeV, yielding an asymmetry\cite{ghu}
\begin{equation}
A_L(\vec{{\rm p}}\alpha; 46\,\,{\rm MeV})=(-3.3 \pm 0.9)\times 10^{-7}.
 \end{equation}
\item[e)] A NIST study of the spin rotation of transversely polarized neutrons in passage through a ${}^4$He target has yielded the upper bound \cite{sno}
\begin{equation}
{d\phi^{{\rm n}\alpha}\over dz}=(1.7\pm 9.1\pm 1.4)\times 10^{-7} {\rm rad/m}.
\end{equation}
(As this observable is the isospin mirror of $\vec{\mathrm{p}}$+$^4$He, it would allow an isoscalar/isovector separation to be done for A=5.)
\item[f)] An upper bound also has been established \cite{nagle2} on the analyzing power for longitudinally polarized protons scattering on deuterium
\begin{equation}
A_L(\vec{{\rm p}} {\rm d}; 15\,\,{\rm MeV}) = (-0.35 \pm 0.85) \times 10^{-7}.
\end{equation}
\end{itemize}

As so few of the attempted experiments in NN and few-body experiments succeeded in isolating nonzero effects at the expected $\sim 10^{-7}$ level,
experimenters turned in the late 1970s and 1980s to more complicated ``two-level" nuclei where chance amplifications of PNC
observables could be exploited.  The origin of this amplification is level mixing:  in the presence of a parity violating interaction described by the Hamiltonian ${\cal H}_w={\cal H}_w^{CC}+{\cal H}_w^{NC}$, the $J^\pi$ strong eigenstates
of same angular momentum and but opposite parity will mix via the weak interaction.  The PNC admixture can be determined from first-order perturbation theory
\begin{eqnarray}
|\psi_{J^+i}>&\simeq&|\phi_{J^+i}>+\sum_k{<\phi_{J^-k}|{\cal H}_w|\phi_{J^+i}>\over E_{J^+i}-E_{J^-k}}|\phi_{J^-k}>\nonumber\\
|\psi_{J^-i}>&\simeq&|\phi_{J^-i}>+\sum_k{<\phi_{J^+k}|{\cal H}_w|\phi_{J^-i}>\over E_{J^-i}-E_{J^+k}}|\phi_{J^+k}>
\label{eq:admix}
\end{eqnarray}
where $i$ and $k$ represent all other quantum labels of the nuclear levels.
For typical nuclear weak matrix elements $<\phi_{J^-k}|{\cal H}_w|\phi_{J^+i}>\sim$ 1 eV, which can be compared to the spacings of neighboring opposite-parity major shells $\sim 10$ MeV.  Consequently the PNC amplitudes  in Eq. (\ref{eq:admix}) are typically$\sim 10^{-7}$.  However, if two states fortuitously form a nearly degenerate parity doublet, the mixing can be significantly enhanced.  In this case, Eq. (\ref{eq:admix}) also simplifies because the two-level mixing will dominate all other contributions.
Many early experiments utilized nuclei with such parity doublets to compensate for experimental sensitivities that
would otherwise have been insufficient to probe ${\cal H}_w$.  

Several experiments yielded nonzero results, and from such
two-level systems one can often extract the magnitude of the nuclear matrix element of ${\cal H}_w$.  But the next step -- deducing from the many-body matrix element constraints on the underlying
weak couplings of ${\cal H}_w$ --  can be problematic due to the uncertainties in model-based calculations of
nuclear wave functions.  Yet there are possible exceptions:
\begin{itemize}
\item[g)] Four independent efforts were mounted to search for the circular polarization of photons emitted in the decay from the $J^P,I=0^-,0$ 1.081 MeV excited state of ${}^{18}$F to the $J^P,I=0^+,0$ ground state.   While rather stringent limits were obtained, the groups failed to find a signal.  The initial state mixes with the nearby $J^P,I=0^+,1$ level at 1.042 MeV, so that $\Delta E=39$ KeV.  Thus this system isolates
the $\Delta I =1$ PNC mixing of greatest interest. The results from the four experiments are in good agreement\cite{ghk,ghl,ghm,ghn}
    \begin{equation}
    P_\gamma=\left\{\begin{array}{ll}
    (-7\pm 20)\times 10^{-4}&{\rm CalTech/Seattle}\\
    (3\pm 6)\times 10^{-4}&{\rm Florence}\\
    (-10\pm 18)\times 10^{-4}&{\rm Mainz}\\
    (2\pm 6)\times 10^{-4}&{\rm Queens}
    \end{array}\right.
    \end{equation}
The experiments reached past the DDH best-value sensitivity while failing to detect a nonzero result.
Nuclear physics uncertainties in this system are unusually modest due to a relationship \cite{hax81} between the mixing
matrix element for the long-range pion-exchange contribution to PNC and the measured axial-charge 
$\beta$-decay rate for the transition from the $J^P,I M_I = 0^+,11$ ground-state of $^{18}$Ne, the isobaric analog of
the 1.042 MeV state in $^{18}$F,  to the 1.081 MeV state in $^{18}$F.

\item[h)] The photon asymmetry in the radiative decay of the polarized $J^P,I={1\over 2}^-,{1\over 2}$ 110 keV first excited state of ${}^{19}$F to the $J^P,I={1\over 2}^+,{1\over 2}$ ground state was measured in two independent experiments.  The mixing here is between these two states, so $\Delta E=110$ keV.  
The measured asymmetries are\cite{gho,ghp}
    \begin{equation}
    A_\gamma=\left\{\begin{array}{ll}
    (-8.5\pm 2.6)\times 10^{-5}&{\rm Seattle}\\
    (-6.8\pm 1.8)\times 10^{-5}&{\rm Mainz}
    \end{array}\right.
    \end{equation}
This mixing matrix element is a sum of isoscalar and isovector amplitudes.  The isovector contribution can again
be related to the measured analog axial-charge $\beta$ decay rate of the $^{19}$Ne ground state to the 110 keV
excited state in $^{19}$F.    The isoscalar contribution
is assumed to scale similarly.  While there are arguments to support this procedure, the axial-charge beta decay
argument is on less firm ground for $^{19}$F than for $^{18}$F.

\item[i)] Two independent experiments looked for circular polarization of photons emitted from the radiative decay of the $J^P,I={1\over 2}^-,{1\over 2}$ 2.789 MeV excited state of ${}^{21}$Ne to the $J^P,I={3\over 2}^+,{1\over 2}$ ground state.  The mixing here is with the nearby 2.795 MeV $J^P,I={1\over 2}^+,{1\over 2}$ excited state with $\Delta E=5.7$ KeV.  The measured circular polarizations are\cite{ghq,ghr}
    \begin{equation}
    P_\gamma=\left\{\begin{array}{ll}
    (24\pm 24)\times 10^{-4}&{\rm Seattle/ChalkRiver}\\
    (3\pm 16)\times 10^{-4}&{\rm ChalkRiver/Seattle}
    \end{array}\right.
    \end{equation}
While this nucleus is sufficiently light that rather sophisticated shell-model calculations can be performed, there
are indirect arguments based on discrepancies between these calculations and measured E1 transitions to/from
the states of interest that nuclear structure uncertainties are not under control \cite{rev1}.  For this reason results
from $^{21}$Ne are frequently omitted from global analyses of hadronic PNC.
\end{itemize}

Finally, a class of novel experiments in heavy atoms has been included in some recent analyses of PNC experiments.   The nuclear anapole moment, a weak radiative correction,
generically becomes the dominant V(electron)-A(nuclear) coupling in heavy atoms due to its growth with
the mass number ($\propto A^{2/3}$) and to the relative weakness of the competing tree-level contribution from direct $Z_0$
exchange \cite{zeld,khrip}.   The resulting electron-nucleus interaction has the form
\begin{equation}
H_w={G_F\over \sqrt{2}}\kappa_{tot}~\vec{\alpha_e}\cdot\vec{J}~\rho(r)
\end{equation}
where $\vec{J}$ and $\rho(r)$ denote the nuclear spin and density and $\vec{\alpha}_e$ is the usual Dirac operator
for atomic electrons, and where the anapole moment dominates $\kappa_{tot}$
in heavy systems, because of the $A^{2/3}$ growth in its contribution to $\kappa_{tot}$.  
Nuclear anapole moments can be isolated through the 
dependence of the atomic PNC signal on nuclear spin (e.g., by examining
variations in PNC signals for different hyperfine levels).   As the largest
contribution to nuclear anapole moments comes from
PNC ground-state parity admixtures generated by ${\cal H}_w$, in principle they provide another test of
the hadronic weak interaction.  Anapole results include
\begin{itemize}
\item[j)] limits on the moment of ${}^{205}$Tl\cite{ghv,ghw},
\begin{equation}
\kappa_{tot}({}^{205}{\rm Tl})=\left\{\begin{array}{ll}
0.29\pm 0.40&{\rm Seattle}\\
-0.08\pm 0.40&{\rm Oxford}
\end{array}\right. ,
\end{equation}
\item[k)] and a measurement of the moment of ${}^{133}$Cs\cite{ghx},
\begin{equation}
\kappa_{tot}({}^{133}\mathrm{Cs})=0.112\pm 0.016\,\,\,{\rm Boulder}.
\end{equation}
\end{itemize}
A rather complex nuclear polarizability arises in the theoretical treatment of anapole moments \cite{hw,fm,hlrm}.  As in the
case of the heavy-nucleus results of Eqs. (\ref{eq:xz}) and (\ref{eq:xw}), the calculations are based on
nuclear models that can only partially capture the relevant physics. 
The accuracy of the constraints on the NN
PNC interaction derived from the experiments is consequently somewhat difficult to assess.

\section{Meson-Exchange Approach}

For the past thirty years the most common theoretical approach in analyzing PNC experiments has been that based on the  meson-exchange model of Desplanques, Donoghue, and Holstein (DDH), developed in 1980\cite{ddh}.  DDH constructed a PNC potential based on $\pi^\pm$, $\rho$, and $\omega$ exchanges with strong vertices described by the Hamiltonian
\begin{eqnarray}
{\cal H}_{\rm st}&=&ig_{\pi NN}\bar{N}\gamma_5\vec{\tau}\cdot\vec{\pi} N
+g_\rho\bar{N}\left(\gamma_\mu+i{\chi_V\over
2M}\sigma_{\mu\nu}k^\nu\right) \vec{\tau}\cdot\vec{\rho}^{\,\mu} N\\
&+&g_\omega\bar{N}\left(\gamma_\mu+i{\chi_S\over
2M}\sigma_{\mu\nu}k^\nu \right)\omega^\mu N\label{eq:bg}
\end{eqnarray}
They chose for their strong couplings $g_{\pi
NN}^2/4\pi\sim 14.4$ and $g_\rho^2/ 4\pi={1\over
9}g_\omega^2/4\pi\sim 0.62$ and, using vector dominance
to connect with the electromagnetic interaction,
$\chi_V=\kappa_p-\kappa_n=3.70$ and
$\chi_S=\kappa_p+\kappa_n=-0.12$.

The weak vertices are based on a Hamiltonian with 
seven phenomenological couplings
\begin{eqnarray}
&&{\cal H}_{\rm wk} ={h^1_\pi\over
\sqrt{2}}\bar{N}(\vec{\tau}\times\vec{\pi})_zN\nonumber\\
&+&\bar{N}\left(h_\rho^0\vec{\tau}\cdot\vec{\rho}^{\,\mu} +h_\rho^1\rho_z^\mu
+{h_\rho^2\over 2\sqrt{6}}(3\tau_z\rho_z^\mu
-\vec{\tau}\cdot\vec{\rho}^{\,\mu})\right)
\gamma_\mu\gamma_5N\nonumber\\
&+&\bar{N}
\left(h_\omega^0\omega^\mu+h_\omega^1\tau_z\omega^\mu\right)\gamma_\mu\gamma_5N
-{h_\rho^1}'\bar{N}(\vec{\tau}\times\vec{\rho}^{\,\mu})_z{\sigma_{\mu\nu}k^\nu\over
2m_N} \gamma_5N.\nonumber\\
\quad\label{eq:vg}
\end{eqnarray}
Couplings to the neutral pseudoscalar
mesons $\pi^0,\,\eta^0,\,{\eta^0}'$ are absent due to Barton's theorem, which requires such couplings to be
CP violating\cite{btn}.  Combining Eqs. (\ref{eq:bg}) and (\ref{eq:vg}) and doing a Fourier transform yields
the coordinate-space DDH potential, 
\begin{eqnarray}
&&V^{\rm PNC}_{DDH}(\vec{r}) =i{h^1_\pi g_{\pi NN}\over
\sqrt{2}}\left({\vec{\tau}_1\times\vec{\tau}_2\over 2}\right)_z
(\vec{\sigma}_1+\vec{\sigma}_2)\cdot
\left[{{\vec p}_1-{\vec p}_2\over 2m_N},w_\pi (r)\right]\nonumber\\
&&-g_\rho\left(h_\rho^0\vec{\tau}_1\cdot\vec{\tau}_2+h_\rho^1\left({\vec{\tau}_1+\vec{\tau}_2\over
2} \right)_z+h_\rho^2{(3\tau_1^z\tau_2^z-\vec{\tau}_1\cdot\vec{\tau}_2)\over
2\sqrt{6}}\right)
\nonumber\\
&&\quad \times\left((\vec{\sigma}_1-\vec{\sigma}_2)\cdot
\left\{{{\vec
p}_1-{\vec p}_2\over 2m_N},w_\rho(r)\right\}\right.\nonumber\\
&+&\left.i(1+\chi_V)\vec{\sigma}_1\times\vec{\sigma}_2\cdot
\left[{{\vec p}_1-{\vec p}_2\over 2m_N},w_\rho
(r)\right]\right)\nonumber\\
&&-g_\omega\left(h_\omega^0+h_\omega^1\left({\vec{\tau}_1+\vec{\tau}_2\over
2}\right)_z
\right)\nonumber\\
&&\quad\times\left((\vec{\sigma}_1-\vec{\sigma}_2)\cdot
\left\{{{\vec
p}_1-{\vec p}_2\over 2m_N},w_\omega (r)\right\}\right.\nonumber\\
&+&\left.i(1+\chi_S)\vec{\sigma}_1\times\vec{\sigma}_2\cdot
\left[{{\vec p}_1-{\vec p}_2\over
2m_N},w_\omega(r)\right]\right)\nonumber\\
&&+
\left({\vec{\tau}_1-\vec{\tau}_2\over 2}\right)_z
(\vec{\sigma}_1+\vec{\sigma}_2)\cdot \left( g_\rho h_\rho^1 \left\{{{\vec p}_1-{\vec
p}_2\over 2m_N},w_\rho(r)\right\}-g_\omega h_\omega^1\left\{{{\vec p}_1-{\vec
p}_2\over 2m_N},w_\omega(r)\right\} \right)
\nonumber\\
&&-g_\rho {h_\rho^1}'i\left({\vec{\tau}_1\times\vec{\tau}_2\over 2}\right)_z
(\vec{\sigma}_1+\vec{\sigma}_2)\cdot \left[{{\vec p}_1-{\vec
p}_2\over 2m_N},w_\rho(r)\right],\nonumber\\
\quad
\label{eq:DDH}
\end{eqnarray}
where $w_i(r)=\exp (-m_ir)/4\pi r$ is the usual Yukawa potential,
$r=|{\vec r}_1 - {\vec r}_2|$ is the relative NN coordinate,
and ${\vec p}_i=-i{\vec\nabla}_i$.  The resulting PNC NN interaction
is described in terms of seven phenomenological weak NN-meson couplings---$h^1_\pi,\,h_V^n,\,{h_\rho^1}'$---though the
constant ${h_\rho^1}'$ is generally not included, since it is a short ranged piece of the dominant pion coupling
and a simple bag model estimate has shown that it is small\cite{pkl}.  This potential
is very closely related to model-independent, threshold $S-P$ interactions,
as discussed below.

The results of PNC analyses are often presented as in Fig. \ref{fig:limits}, in terms of
constraints on the weak couplings.  However, the weak NN amplitudes depend
on a product of weak and strong couplings, so that comparisons of extracted weak
couplings can become problematic if experiments are not analyzed with a common
set of strong coefficients.  We will return to this point in later discussions.

Using quark model and symmetry methods, DDH attempted to calculate values for these weak parameters.
However, because of strong interaction uncertainties, an accurate estimate proved impossible, so DDH
instead quoted reasonable ranges and ``best values" -- or perhaps more accurately, ``best guesses" -- for each.  These are listed in Table 1 together with values
estimated by some other investigators.  One issue relevant to later discussions is that the DDH
best value for $h_\pi^1$ is higher than found by some others.   Skyrme and quark-soliton
models tend to give smaller values for $h_\pi^1$ \cite{meissner,lhk}.  A large-$N_c$
result has also been obtained recently \cite{zhun}.
\begin{table}
\begin{center}
\begin{tabular}{|c|c|c|c|c|}
\hline \quad   & DDH\cite{ddh} & DDH\cite{ddh} & DZ\cite{dze} &
FCDH\cite{fcd}\\
Coupling & Reasonable Range & ``Best" Value &  &  \\ \hline
$h^1_\pi$ & $0\rightarrow 30$ &+12 &+3&+7\\
$h_\rho^0$& $30\rightarrow -81$&$-30$&$-22$&$-10$\\
$h_\rho^1$& $-1\rightarrow 0$& $-0.5$&+1&$-1$\\
$h_\rho^2$& $-20\rightarrow -29$&$-25$&$-18$&$-18$\\
$h_\omega^0$&$15\rightarrow -27$&$-5$&$-10$&$-13$\\
$h_\omega^1$&$-5\rightarrow -2$&$-3$&$-6$&$-6$\\ \hline
\end{tabular}
\caption{Weak NN-meson couplings as calculated in Refs.
\cite{ddh,dze,fcd}, in units
of $G_F F_\pi^2 /2 \sqrt{2} \sim 0.38 \times 10^{-7}$, where $G_F$ is Fermi's weak
coupling and $F_\pi$  the pion decay constant.}
\end{center}
\label{tab0}
\end{table}

Analysis of key experiments described in the previous section in terms of these
unknown weak couplings has produced the well known graph \cite{hw} shown in Fig. \ref{fig:limits},
which exploits the fact that certain isoscalar and isovector combinations of the DDH parameters
dominate the experimental observables.  This plot dates from 2001 and as we will show
later, should be modified to reflect more recent developments in PNC studies.

\begin{figure}
\begin{center}
\epsfig{file=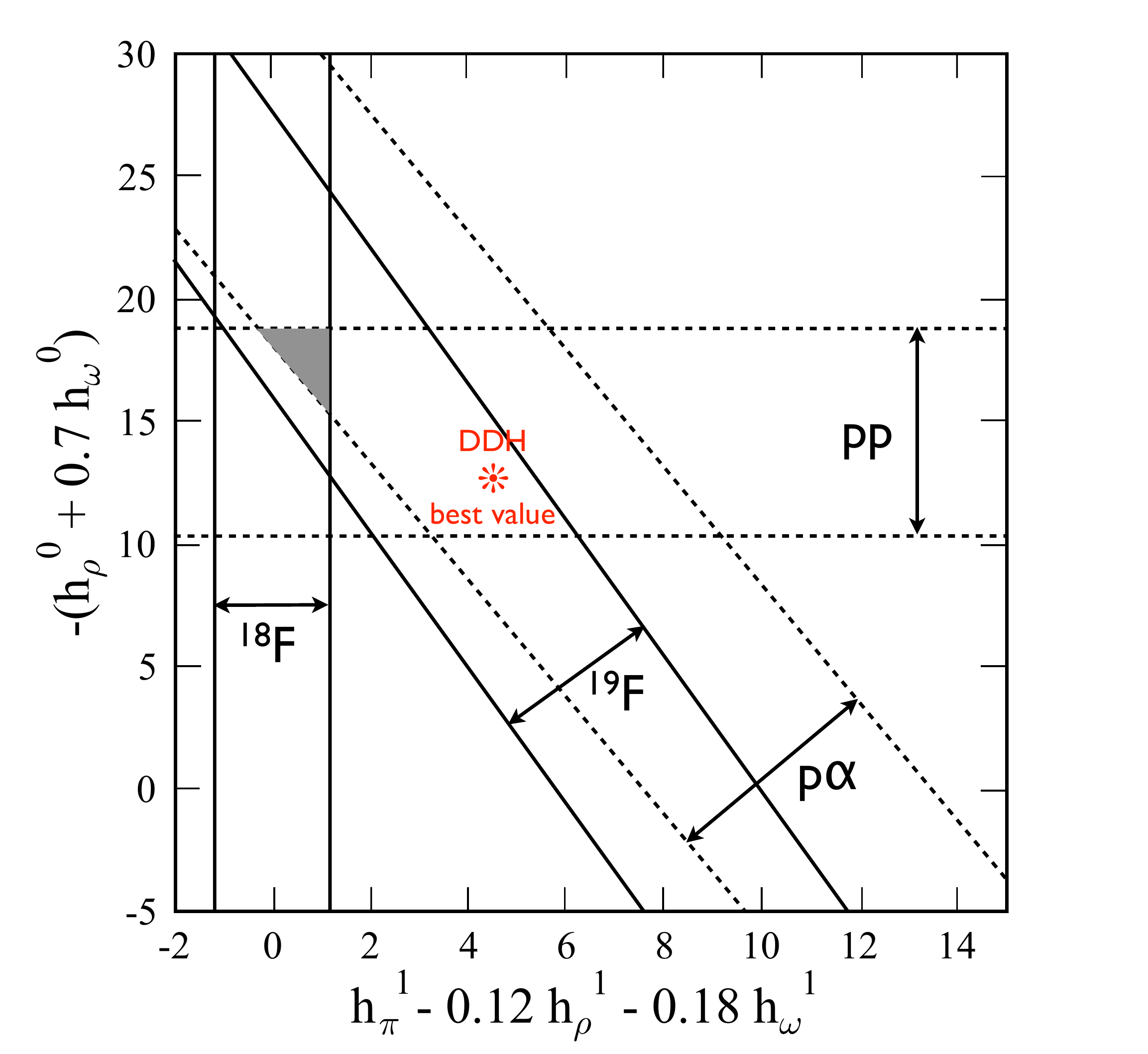,width=13cm}
\caption{Experimental constraints on linear combinations of isoscalar and isovector DDH couplings
(in units of 10$^{-7}$), taken from
the 2001 work of \cite{hw}, displaying bounds from four experiments where it is believed
that theoretical analysis uncertainties are under reasonable control: pp, p$\alpha$,
$^{18}$F, and $^{19}$F.  The small shaded triangle is consistent with all four experiments.  The
DDH best value point is also shown. Later we show that the data on $\vec{p}+p$ subsequently obtained at
TRIUMF \cite{TRIUMF} and the analysis of
Ref. \cite{carlson} have a significant impact on this plot.}
\label{fig:limits}
\end{center}
\end{figure}

\newpage
For the reasons described previously, the most reliable constraints on PNC come from a limited number
of measurements  where PNC has been seen or sharply limited, and where the analysis of the observables
can be done reliably either with {\it ab initio} techniques or by relating observables to other measured
quantities such as axial-charge beta decay rates.  
Figure \ref{fig:limits} -- a figure that has been used frequently over the past decade, despite its 2001
vintage \cite{hw} -- shows that the allowed bands for these observables -- pp, p$\alpha$, $^{18}$F, and 
$^{19}$F --  intersect in a limited region defined by linear combinations of  isoscalar and isovector weak
couplings.  The resulting value of the isoscalar
coupling $-(h_\rho^0+0.7h_\rho^0) \sim (15-18) \times 10^{-7}$ can be compared to the DDH best value
$\sim 12.5 \times 10^{-7}$ and reasonable range of  (-15.4 $\leftrightarrow$ 39.0) $\times 10^7$.
The two alternative sets of theoretical weak couplings given in Table \ref{tab0} yield
$\sim 11 \times 10^{-7}$\cite{dze} and $\sim 7.3 \times 10^{-7}$\cite{fcd}.  
%

If we consider the extended region where at least three of the four bands overlap, this range
would expand to $\sim (13-25) \times 10^{-7}$.  Thus the conclusion from
the 2001 analysis is that the experimental value
is likely comparable to or somewhat larger than the DDH best value, though quite consistent
with the reasonable range.
One also sees that there is some tension between the isoscalar coupling derived
from  $\vec{p}+p$ asymmetry and that derived from other observables:
the $\vec{p} + p$ asymmetry tends to favor somewhat weaker isoscalar couplings.  This tension
will be re-examined in the next section, in light of results on $\vec{p}+p$ obtained after the
analysis of \cite{hw} that led to Fig. \ref{fig:limits}.   In deriving the $\vec{p}+p$ band,
the experimental result was  ``marginalized" over degrees of freedom 
not shown in the plot, resulting in a broadening of the band: 
this includes the isotensor contribution to $\vec{p}+p$ as well as a small
isoscalar contribution orthogonal to the combination $-(h_\rho^0+0.7h_\rho^0)$.  New results
discussed in the next section allow improvements in this process.

The discrepancy in the case of the long-range isovector pion coupling $h_\pi^1$ is much more
substantial.  The range of values consistent with the combined results of the four ${}^{18}$F experiments,
$P_\gamma = (1.2 \pm 3.9) \times 10^{-4}$\cite{ghn}, and the relationship established by measured $\beta$
decay rates with  the inclusion of all identified theory uncertainties, $|P_\gamma| = (4.2 \pm 1.0) \times 10^3~ |h^1_\pi|$ \cite{gho}, yields
\begin{equation}
|h^1_\pi| \lesssim 1.3 \times 10^{-7}.
 \end{equation}
Thus $|h^1_\pi|$ is
substantially smaller than the DDH best value estimate,  $4.6 \times 10^{-7}$.
Although ${}^{18}$F is a complex nucleus, it
 is difficult to attribute the discrepancy to nuclear structure
uncertainties because the PNC mixing matrix element needed in the analysis can
be determined from
the analog beta decay of ${}^{18}$Ne\cite{rev1,hax81}:  with this constraint, the residual model dependence --
that is, the variation that can be achieved by using any of the wide range of $^{18}$F wave functions to relate
$^{18}$F PNC to $^{18}$N $\beta$ decay -- was found to be
$\lesssim 14\%$ \cite{rev1}.  The axial-charge
$\beta$ decay argument relates bare couplings, and thus eliminates concerns about operator renormalization
that accompanies shell-model calculations. 

This $h_\pi^1$ result is supported by two others.  A lattice QCD calculation of $h^1_\pi$ was recently performed\cite{wasem}
for a lattice size $L\sim$2.5 fm, lattice spacing $a_s\sim0.123$ fm, and a pion mass $m_\pi \sim$ 389 MeV,
accomplishing a goal envisioned some time ago\cite{bs}.
The result obtained
\begin{equation}
h^1_\pi(\mathrm{connected)} = (1.099 \pm 0.505~ \mathrm{(stat)}~ {}^{+0.508}_{-0.604}~ \mathrm{(sys)}) \times 10^{-7}
\end{equation}
is consistent with the $^{18}$F result and also well below the DDH best value.  This calculation is a first
step: it did not include nonperturbative renormalization of the bare parity-violating operators, a chiral extrapolation
to the physical pion mass, or contributions from disconnected (quark loop) diagrams, though it was argued
that these omitted effects would be expected to lie within the designated systematic error.  Nevertheless,
the result should be considered preliminary, pending future work with physical pions and including disconnected
diagrams

The ${}^{18}$F result is also consistent with the interpretation of a recent measurement of a PNC triton emission
asymmetry coefficient in the reaction ${}^6$Li(n,$\alpha$)$^3$H with polarized 
cold neutrons.  From the nonzero result  obtained, $a_\mathrm{PNC} = (-8.8 \pm 2.1) \times 10^{-8}$\cite{tri}, 
the following limit was obtained
\begin{equation}
|h^1_\pi| \lesssim 1.1 \times 10^{-7}.
\end{equation}
However, we are reluctant to adopt this result as independent evidence for a suppressed $h^1_\pi$ because the 
reaction theory on which the limit is based is rather simple in its use of a cluster decomposition and
schematic interactions.  The experimental result provides good motivation for 
further theoretical work.

\section{The TRIUMF 221 MeV $\vec{p}+p$ Measurement}
A somewhat discouraging aspect of the effort to understand hadronic parity PNC has been the
slow pace of results since the 1980s.   The decade of the 1980s gave us several important
measurements:  the longitudinal asymmetries for $\vec{p}$+p and $\vec{p}+\alpha$, the significant
limits from the Queens and Florence group on $P_\gamma(^{18} \mathrm{F})$ that pointed to a problem with $h^1_\pi$,
our cleanest handle on neutral current contributions to PNC, and the measured $A_\gamma$ for
$^{19}$F.  But the last twenty years have proven more difficult.  A promising LANL effort on
$A_\gamma(\vec{n}p)$ -- an experiment that could confirm conclusions drawn from $^{18}$F --
in the end yielded a limit comparable to that established in the Grenoble effort of 1977.  Another effort
is now underway\cite{SNS} at the SNS cold neutron beamline that hopefully will meet with greater success.  The NIST
experiment on neutron spin rotation in He also provided only an upper bound: had PNC been detected,
a separation of isoscalar and isovector contributions to PNC could have been made by combining
this result with the existing $\vec{p}+^4$He measurement.  As significant difficulties were encountered 
in this experiment, new ideas may be needed before another attempt can be made \cite{snow}.  
The anapole moment of Cs was measured -- a remarkable feat that does establish the expected
sharp growth in this weak radiative correction with increasing A -- but it is difficult to envision any 
confident determination of PNC couplings from the measurement given the complexity of the
nuclear polarizability that governs this moment.  Finally, the recent measurement of PNC in
 ${}^6$Li(n,$\alpha$)$^3$H is a significant advance, but {\it ab initio} methods for handling
 the associated analysis are not yet in hand.
 
 One significant new result is the TRIUMF medium-energy measurement of $\vec{p}+p$,
 reported in 2001 and 2003 \cite{TRIUMF,ramsey},
 \begin{equation}
 A_L(221~\mathrm{MeV})  = (0.84 \pm 0.29 \mathrm{(stat)~}\pm 0.17 \mathrm{(syst)}) \times 10^{-7}. 
 \end{equation}
  This
measurement was performed at an energy where the ${}^1S_0-{}^3P_0$ amplitude
nearly vanishes, and consequently where the ${}^3P_2-{}^1D_2$ 
partial wave dominates the scattering.   We discuss this result in detail here because, as explained
below, this result has been combined with others in a way that has obscured its impact on
Fig. \ref{fig:limits}.

The result, together with the lower energy $\vec{p}+p$ measurements, was analyzed by 
Carlson {\it et al.} \cite{carlson} using  several modern strong potentials - Argonne v18 (AV18) , Bonn-2000, Nijmegen-I.
The full scattering problem with strong, Coulomb, and weak potentials
was solved in both coordinate and momentum space. This work remains state-of-the-art,
the definitive treatment of PNC in $\vec{p}+p$.   The approach resembles early work by Driscoll and
Miller \cite{dm} and confirmed many of their conclusions, including the importance of distorted waves
even at rather modest momentum transfers \cite{dm}.  While channels up to angular momentum J=8
were included,  it was shown that
the region up to the laboratory energy of 221 MeV could be described accurately by
retaining the first two PNC partial waves of the DDH potential, 
${}^1S_0-{}^3P_0$ and  ${}^3P_2-{}^1D_2$.  

The results of Carlson {\it et al.} were presented as bounds on DDH weak couplings, 
and those bounds appears roughly consistent with the pre-TRIUMF analysis shown in Fig. 1.  For
this reason Fig. 1 is frequently used today as a summary of PNC constraints.  However the analysis
of Ref. \cite{carlson} used
CD Bonn strong couplings that differ substantially from those of the DDH potential,
\begin{equation}
{g_\rho^{CD~Bonn} \over g_\rho^{DDH} } \sim 1.16~~~{g_\omega^{CD~Bonn} \over g_\omega^{DDH} } \sim 1.89~~~{\chi_V^{CD~Bonn} \over \chi_V^{DDH}} \sim 1.65~~~{\chi_S^{CD~Bonn} \over \chi_S^{DDH}} =0.
\end{equation}
This, of course, does not affect the validity of the calculations -- only their interpretation when combined
with calculations that may have made other choices in their strong couplings.  The fact that the Carlson 
{\it et al.} obtained numerical values for weak couplings that were roughly compatible with those shown
in Fig. 1 is likely the reason Fig. 1 remains in wide use.  In fact, when the differences in strong couplings
are taken into account, the TRIUMF data and the Carlson {\it et al.}  analysis require an important 
revision of Fig. 1 that brings
the experiments into better agreement, as we explain below.
(We note that the confusion that can occur when comparisons of experiments and calculations are done at the level
of weak couplings is an old issue.  To avoid such confusion,
the authors of \cite{rev1} advocated quoting experimental
constraints in terms of the product of weak and strong couplings, but this suggestion did not catch on.
As pionless effective field theory treatments essentially force comparisons to be made
at the NN amplitude level, not at the weak vertex level, the situation may be
changing,  as we discuss in Sec. 6.)

\begin{figure}
\begin{center}
\epsfig{file=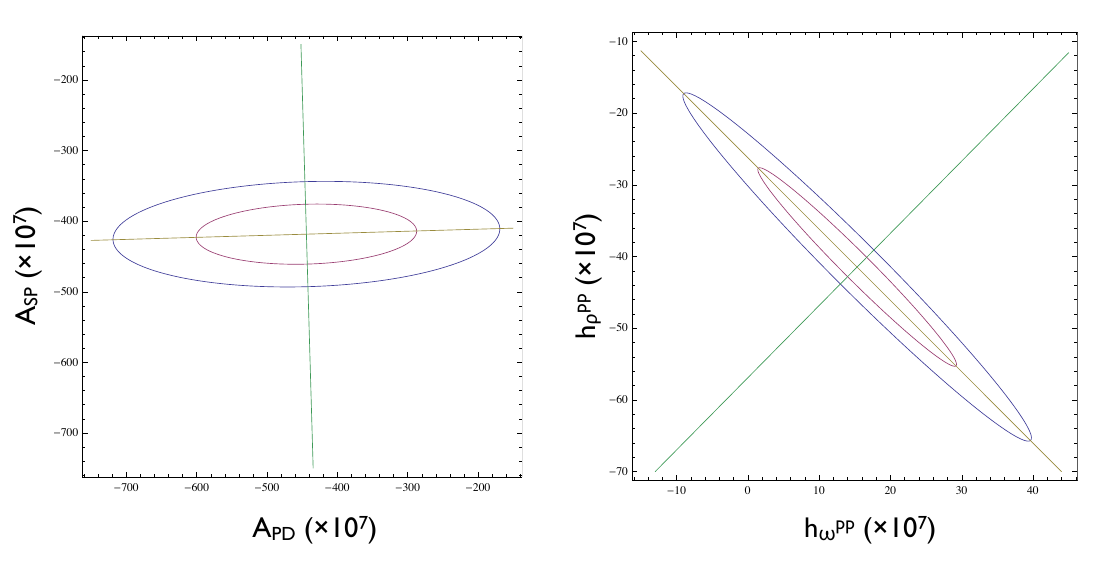,width=13.8cm}
\caption{Left panel: 68\% and 90\% c.l.
constraints on the ${}^1S_0-{}^3P_0$ and ${}^3P_2-{}^1D_2$
coefficients $A_{SP}=h_\rho g_\rho (2 + \chi_V) + h_\omega g_\omega (2 + \chi_S)$ and
$A_{PD}=h_\rho g_\rho \chi_V + h_\omega g_\omega \chi_S$ derived from the 13.6, 45, and
221 MeV $\vec{p}+p$ measurements and the calculations of \cite{carlson}.  Right panel:
Constraints on the DDH potential weak couplings imposed by the
results on the left when interpreted in terms of DDH strong couplings.}
\label{fig:ellipses}
\end{center}
\end{figure}

The overall strengths of the $\rho$ and $\omega$ couplings that enter into $\vec{p}+p$ scattering
depend on the isospin combinations
\begin{equation}
g_\rho h_\rho^{pp} =g_\rho( h_\rho^0 + h_\rho^1 + {h_\rho^2 \over \sqrt{6}})~~~~~h_\omega^{pp} = g_\omega (h_\omega^0 + h_\omega^1)
\end{equation}
These parameters appear in the DDH potential alone as well as in the combinations
\begin{equation}
g_\rho h_\rho^{pp} \chi_V~~~~~g_\omega h_\omega^{pp} \chi_S.
\end{equation}
Thus the parameter space has four degrees of freedom.  However in the plane-wave Born approximation this
collapses to two degrees of freedom associated with the coupling combinations
\begin{equation}
A_{SP} \equiv g_\rho h_\rho^{pp} (2 + \chi_V) + g_\omega h_\omega^{pp} (2 +\chi_S)~~~~~~A_{PD} \equiv g_\rho h_\rho^{pp} \chi_V + g_\omega h_\omega^{pp} \chi_S,
\end{equation}
the coefficients of the ${}^1S_0-{}^3P_0$ and  ${}^3P_2-{}^1D_2$ amplitudes.  Thus in this limit
it is possible to define strictly equivalent sets of weak couplings for arbitrary variations in the
strong couplings ${g_\rho, \chi_V, g_\omega, \chi_S}$.  This appears to be the procedure Carlson {\it et al.}
used in defining their DDH-equivalent and adjusted-DDH couplings.  It is also the approximation
we make here.

We use the ${}^1S_0-{}^3P_0$ (J=0) and 
${}^1S_0-{}^3P_0$+${}^3P_2-{}^1D_2$ (J=0+2) results of \cite{carlson} which are given separately in the paper.  (The
J=0+2 result is virtually identical to the full result that includes all angular momentum channels though J=8,
showing that only two channels are important.)
Treating the experimental data as described in \cite{carlson} and fitting the 
three experimental data points, we find the constraints on $A_{SP}$ and $A_{PD}$ displayed in
Fig. \ref{fig:ellipses}.  This procedure thus gives us constraints on the product of weak and strong couplings,
the quantities directly determined by experiment.  The procedure was cross-checked by
plugging in CD-Bonn strong couplings and generating
the analog of Fig. 8 of \cite{carlson} (the $\chi^2$ ellipses  for $h_\rho^{pp}$ and
$h_\omega^{pp}$ when CD-Bonn strong couplings
are adopted).  The agreement is quite good.   We can then derive the similar constraints on $h_\rho^{pp}$ and
$h_\omega^{pp}$ for DDH strong couplings, shown in the right panel of Fig. \ref{fig:ellipses}.  
The 68\% and 90\% confidence level contours are based on the 
appropriate $\chi^2$ for three data points and two parameters (one degree of freedom).

The semi-major and semi-minor axes of the ellipse provide the orthogonal constraints
\begin{eqnarray}
0.710 h_\rho^{pp~DDH}+0.705 h_\omega^{pp~DDH} &=& -18.63 \pm 1.90 \nonumber \\
0.705 h_\rho^{pp~DDH} - 0.710 h_\omega^{pp~DDH} &=& -40.12 \pm 19.55
\end{eqnarray}
at 68\% c.l.  While the first constraint is close to that needed for a revised Fig. \ref{fig:limits}, we can combine
both results to limit the quantity of interest for a revised Fig. \ref{fig:limits}.  That is, because of the 221 MeV
TRIUMF result, we can bound any needed  linear combination of $h_\rho^{pp}$ and $h_\omega^{pp}$. From
the axis rotation illustrated in Fig. \ref{fig:limitsII} we find
\begin{equation}
h_\rho^{pp~DDH}+0.7 h_\omega^{pp~DDH} = -30.75 \pm 4.75
\end{equation} 
Allowing the small isovector and larger isotensor contributions to this quantity to vary arbitrarily throughout
the DDH reasonable ranges, one finds
\begin{equation}
-(h_\rho^{0~DDH}+0.7 h_\omega^{0~DDH}) =  25.9^{+6.0}_{-6.1},
\label{eq:finalDDH}
\end{equation} 
a result that can then be placed on the revised limits graph shown in Fig. \ref{fig:limitsII}.

The new analysis removes the tension we observed in
Fig. \ref{fig:limits}.   The band of values for  $-(h_\rho^{0~DDH}+0.7 h_\omega^{0~DDH})$ consistent
with the constraints from $\vec{p}+\alpha$, $^{19}$F, and $^{18}$F, $\sim (15-31) \times 10^7$ is in excellent
agreement with that obtained from $\vec{p}+p$ in Eq. (\ref{eq:finalDDH}).  The region allowed by
all four experiments, $\sim (20-31) \times 10^7$, is centered on a value that is approximately twice
the DDH best value.

\begin{figure}
\begin{center}
\epsfig{file=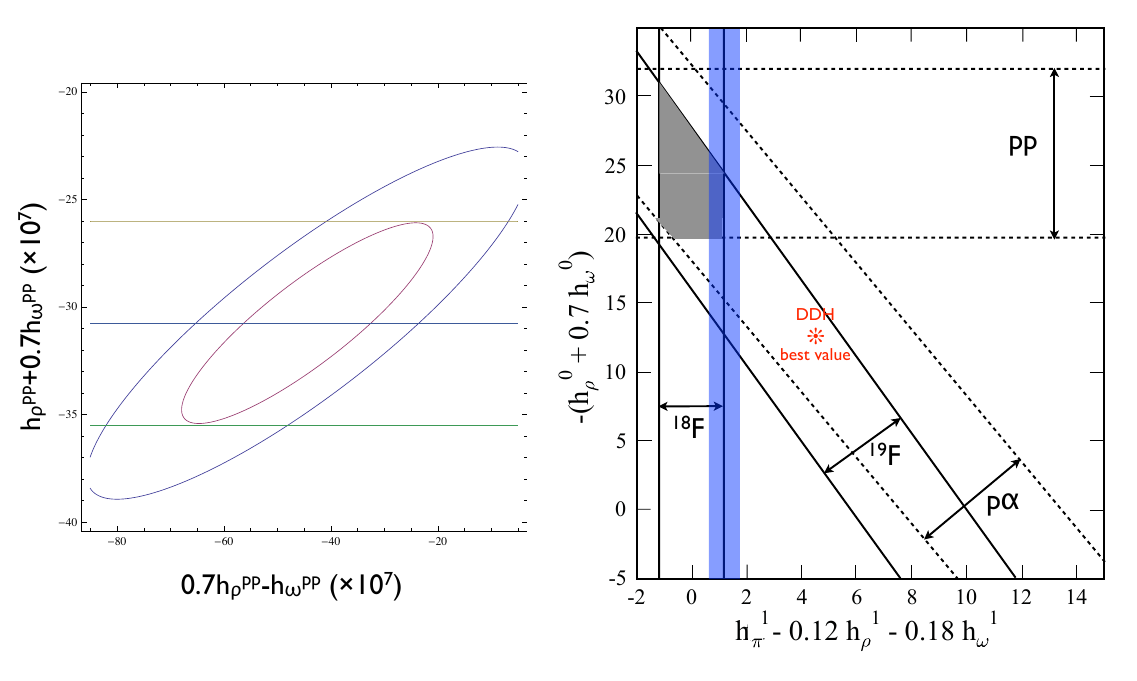,width=13.9cm}
\caption{Left panel: The 68\% and 90\% c.l. constraints on $h_\rho^{pp}$ and $h_\omega^{pp}$ 
from the right panel of Fig. \ref{fig:ellipses} are rotated into a form
that can be use in redrawing Fig. \ref{fig:limits}.  Right panel:   Our suggested replacement
for Fig \ref{fig:limits} using the new analysis of $\vec{p}+p$ to constraint the isoscalar
combination $h_\rho^{0} +0.7 h_\omega^0$ and showing
a recent lattice QCD estimate  of the connected-diagram
contribution to $h^1_\pi$ (blue vertical band).  Units are 10$^{-7}$. The gray shaded area is 
consistent with all experiments.}
\label{fig:limitsII}
\end{center}
\end{figure}

Figure \ref{fig:limitsII} also includes Wasem's recent lattice QCD calculation \cite{wasem} of $h_\pi^1$, shown as the blue
vertical band.  We have plotted this result by adding the systematic error to the statistical error, and note that it is difficult 
to estimate the effects of the omitted disconnected diagrams and of
the nonphysical pion mass employed (389 MeV).   

Even without new experiments, there are important steps that can be taken to further tighten the bounds
shown in Fig. \ref{fig:limitsII}.  First, Wasem's benchmark calculation has demonstrated that lattice QCD has
reached the point that it can impact hadronic PNC analyses.  The completion of the $h_\pi^1$ 
work -- the evaluation of the disconnected contribution and the repetition of the connected
calculation with physical or near-physical pion masses -- appears to be feasible with computing resources
now coming available (though the requirements may be 10-100 times the CPU investment made in the
connected calculation, where the contractions required about 6 months of running
on LLNL's Edge GPU cluster).  There is also a second important opportunity:  the $\vec{p}+p$ band
shown in Fig. \ref{fig:limitsII} requires one to remove the isovector and isotensor contributions from
$h^{pp}_\rho + 0.7 h_\omega^{pp}$.  This was done by allowing these contributions to vary over their
full DDH reasonable ranges, resulting in a substantial expansion of the uncertainty on
$h^{0}_\rho + 0.7 h_\omega^{0}$.  As the isovector couplings are thought to be very small, 
this step is controlled by the DDH reasonable-range uncertainty on $h^{2}_\rho$.  This parameter has
no disconnected contribution in lattice QCD and thus should be calculable to high accuracy.   The
completion of a lattice QCD calculation of $h_\rho^2$ could thus significantly narrow the
$\vec{p}+p$ band in Fig. \ref{fig:limitsII}.  The CalLat Collaboration has proposed a program
of lattice QCD calculations focused on $h^1_\pi$ and $h^2_\rho$ \cite{CalLat}.

Second, we would advocated that the authors of \cite{carlson} revisit that important work.  It should be
possible to express the $\vec{p}+p$ asymmetries calculated there analytically, as coefficients associated
with combinations of weak and strong parameters, $h_\rho g_\rho$, $h_\rho g_\rho \chi_V$,
$h_\omega g_\omega$, and $h_\omega g_\omega \chi_S$, evaluated for each of the three energies
of interest (13.6, 45, and 221 MeV).  The coefficients could be tabulated for the
three strong potentials explored in \cite{carlson} and for a range of reasonable single-nucleon form-factor 
masses.  (A form factor was used in \cite{carlson} because of the large relative momentum transfer
at $E_\mathrm{lab} = 221$ MeV.)  This would allow one to simplify the analysis described above --
and guarantee its correspondence with the exact numerical results.  The dependence of the expansion coefficients
on the choice of strong potential and on the value of the form-factor mass would provide a good measure
of strong-interaction uncertainties.

\section{The DDH Potential and its $S-P$ Reduction}
At sufficiently low energies in elementary two-nucleon systems the matrix elements of the PNC interaction
can be written in terms of five elementary SP amplitudes, as Danilov emphasized in early work.
Because long-range pion exchange contributes to the $^3S_1-^3P_1$ channel, the short-range structure 
can be more readily distinguished in this amplitude, through the contribution of higher partial waves.
Thus the zero-range approximation to the low-energy PNC interaction is sometimes elaborated through the
addition of a sixth parameter, the pion range.

As recent treatments based on lowest-order effective field theories (EFTs) begin with
the threshold behavior of PNC amplitudes, here we take a similar approach with
the DDH potential, to illustrate the compatibility of the potential and EFT approaches.  By doing
a Taylor expansion on the initial and final states, the DDH potential can be 
expanded in a power series in derivatives compared to $m$, where $m$ represents $m_\pi$, $m_\rho$, or $m_\omega$,
while the DDH Yukawa potential effectively contracts to a contact form.
The Danilov limit, retention of the lowest order contribution corresponding to S$\leftrightarrow$P 
amplitudes, corresponds to the first term in the expansion where the identification
\[ {e^{-mr} \over 4 \pi r} \leftrightarrow {1 \over m^2} \delta(\vec{r}) \]
is made in Eq. (\ref{eq:DDH}).  [Whether one considers the delta function written above as a true delta
function or the representation of a ``fuzzy" one, $m^2 e^{-mr}/4 \pi r$, is really a matter of taste \cite{Lepage},
in that PNC will always be treated perturbatively.]

A bit of work is needed in expanding the DDH potential to lowest order
because ten operators arise in the  reduction of Eq. (\ref{eq:DDH}), 
but only five of these are independent in the S-P limit.  Defining
\[ \stackrel{\textstyle \leftrightarrow}{\nabla}_S \delta(\vec{r}) \equiv  \stackrel{\textstyle \leftarrow}{\nabla} \delta(\vec{r})+ \delta(\vec{r})\stackrel{\textstyle \rightarrow}{\nabla} ~~~~~~~~\stackrel{\textstyle \leftrightarrow}{\nabla}_A \delta(\vec{r}) \equiv - \stackrel{\textstyle \leftarrow}{\nabla} \delta(\vec{r})+ \delta(\vec{r})\stackrel{\textstyle \rightarrow}{\nabla} \]
one finds the following identities (valid in S-P matrix elements)
\begin{eqnarray}
\stackrel{\textstyle \leftrightarrow}{\nabla}_A \delta(\vec{r}) \cdot (\vec{\sigma}_1-\vec{\sigma}_2) \vec{\tau}_1 \cdot \vec{\tau}_2 & \equiv &- \stackrel{\textstyle \leftrightarrow}{\nabla}_A \delta(\vec{r}) \cdot (\vec{\sigma}_1-\vec{\sigma}_2)-2 i (\vec{\sigma}_1 \times \vec{\sigma}_2) \cdot \stackrel{\textstyle \leftrightarrow}{\nabla}_S \delta(\vec{r}) \nonumber \\
\stackrel{\textstyle \leftrightarrow}{\nabla}_S \delta(\vec{r}) \cdot i(\vec{\sigma}_1 \times \vec{\sigma}_2) \vec{\tau}_1 \cdot \vec{\tau}_2 &\equiv& -2 \stackrel{\textstyle \leftrightarrow}{\nabla}_A \delta(\vec{r}) \cdot (\vec{\sigma}_1-\vec{\sigma}_2)- i (\vec{\sigma}_1 \times \vec{\sigma}_2) \cdot \stackrel{\textstyle \leftrightarrow}{\nabla}_S \delta(\vec{r}) \nonumber \\
\stackrel{\textstyle \leftrightarrow}{\nabla}_A \delta(\vec{r}) \cdot (\vec{\sigma}_1 - \vec{\sigma}_2) (\tau^z_1 + \tau^z_2) &\equiv& - \stackrel{\textstyle \leftrightarrow}{\nabla}_S \delta(\vec{r}) \cdot i (\vec{\sigma}_1\times \vec{\sigma}_2)(\tau^z_1 + \tau^z_2) \nonumber \\
\stackrel{\textstyle \leftrightarrow}{\nabla}_A \delta(\vec{r}) \cdot (\vec{\sigma}_1 + \vec{\sigma}_2) (\tau^z_1 - \tau^z_2) &\equiv& - \stackrel{\textstyle \leftrightarrow}{\nabla}_S \delta(\vec{r}) \cdot (\vec{\sigma}_1+ \vec{\sigma}_2) i (\vec{\tau}_1 \times \vec{\tau}_2)_z \nonumber \\
\stackrel{\textstyle \leftrightarrow}{\nabla}_A \delta(\vec{r}) \cdot (\vec{\sigma}_1 - \vec{\sigma}_2) (\vec{\tau}_1 \otimes \vec{\tau}_2)_{20} &\equiv& - \stackrel{\textstyle \leftrightarrow}{\nabla}_S \delta(\vec{r}) \cdot i(\vec{\sigma}_1\times \vec{\sigma}_2)(\vec{\tau}_1 \otimes \vec{\tau}_2)_{20} 
\label{eq:identity}
\end{eqnarray}
where the rank-two tensor product $(\vec{\tau}_1 \otimes \vec{\tau}_2)_{20} \equiv (3 \tau_1^z \tau_2^z - \vec{\tau}_1 \cdot \vec{\tau}_2)/\sqrt{6}$.
    
These identities can be used to write the most general lowest-order (LO) PNC potential in many equivalent ways.  We
choose to retain the five operators used in \cite{rev1}, using Eqs. (\ref{eq:identity}) to eliminate the remaining five, obtaining
\begin{eqnarray}
V^{PNC}_{LO}(\vec{r}) &=&\Lambda_0^{{}^1S_0-{}^3P_0} \left( {1 \over i} {\stackrel{\textstyle \leftrightarrow}{\nabla}_A \over 2m_N} {\delta(\vec{r}) \over m_\rho^2}  \cdot (\vec{\sigma}_1-\vec{\sigma}_2) - {1 \over i} {\stackrel{\textstyle \leftrightarrow}{\nabla}_S \over 2m_N}  {\delta(\vec{r})\over m_\rho^2} \cdot i(\vec{\sigma}_1 \times\vec{\sigma}_2) \right)\nonumber \\
 &+&\Lambda_0^{{}^3S_1-{}^1P_1} \left( {1 \over i}{ \stackrel{\textstyle \leftrightarrow}{\nabla}_A \over 2m_N} {\delta(\vec{r}) \over m_\rho^2} \cdot (\vec{\sigma}_1-\vec{\sigma}_2) + {1 \over i}  {\stackrel{\textstyle \leftrightarrow}{\nabla}_S \over 2 m_N} {\delta(\vec{r}) \over m_\rho^2} \cdot i(\vec{\sigma}_1 \times\vec{\sigma}_2) \right) \nonumber \\
&+&\Lambda_1^{{}^1S_0-{}^3P_0}  {1 \over i} {\stackrel{\textstyle \leftrightarrow}{\nabla}_A \over 2 m_N} { \delta(\vec{r}) \over m_\rho^2} \cdot (\vec{\sigma}_1 - \vec{\sigma}_2)(\tau_1^z+\tau_2^z) \nonumber \\
 &+& \Lambda_1^{{}^3S_1-{}^3P_1} {1 \over i} {\stackrel{\textstyle \leftrightarrow}{\nabla}_A \over 2m_N} {\delta(\vec{r}) \over m_\rho^2}  \cdot (\vec{\sigma}_1 + \vec{\sigma}_2) (\tau_1^z-  \tau_2^z)  \nonumber \\
&+&\Lambda_2^{{}^1S_0-{}^3P_0}  {1 \over i} {\stackrel{\textstyle \leftrightarrow}{\nabla}_A \over 2 m_N}  {\delta(\vec{r}) \over m_\rho^2} \cdot (\vec{\sigma}_1 -\vec{\sigma}_2)(\vec{\tau}_1 \otimes \vec{\tau}_2)_{20}
\label{eq:LO}
\end{eqnarray}
where the subscripts on the coefficients $\Lambda$ denote the change in isospin $\Delta I$ induced by the operator,
and the superscripts denote the transitions induced by the operators.  (The linear combinations
of the two isoscalar operators formed in the first two
lines above ensure simple projections.) This hermitian S-P PNC potential is the most general
contact interaction that can be constructed, once the identities of Eq. (\ref{eq:identity}) are used to remove
redundant operators in favor of the five we have chosen.   Consequently the $1/m$ expansion of the DDH potential that
keeps only the LO terms must have this form.  This leads to an identification of DDH weak couplings with the 
coefficients of the potential,
\begin{eqnarray}
\Lambda_{0~DDH}^{{}^1S_0-{}^3P_0} &=& - g_\rho h_\rho^0 (2+ \chi_V) - g_\omega h_\omega^0(2+\chi_S) ~~~~~~~~~~~~~~~~~BV~\rightarrow 2.11\cdot 10^{-5} \nonumber \\
\Lambda_{0~DDH}^{{}^3S_1-{}^1P_1} &=& ~~  g_\omega h_\omega^0 \chi_S - 3g_\rho h_\rho^0 \chi_V~~~~~~~~~~~~~~~~~~~~~~~~~~~~~~~BV~ \rightarrow 3.55 \cdot 10^{-5}  \nonumber \\
\Lambda_{1~DDH}^{{}^1S_0-{}^3P_0} &=&   - g_\rho h_\rho^1 (2+ \chi_V) - g_\omega h_\omega^1 (2+\chi_S)~~~~~~~~~~~~~~~~~BV~ \rightarrow 0.21  \cdot 10^{-5}  \nonumber \\
\Lambda_{1~DDH}^{{}^3S_1-{}^3P_1} &=&    {1 \over \sqrt{2}}  g_{\pi NN} h_\pi^1 \left( {m_\rho \over m_\pi} \right)^2 + g_\rho (h_\rho^1-h_\rho^{1\prime})-g_\omega h_\omega^1 ~~~BV~ \rightarrow 13.4 \cdot 10^{-5}\nonumber \\
\Lambda_{2~DDH}^{{}^1S_0-{}^3P_0} &=& - g_\rho h_\rho^2 (2+ \chi_V)~~~~~~~~~~~~~~~~~~~~~~~~~~~~~~~~~~~~~~BV~ \rightarrow 1.52 \cdot 10^{-5} \nonumber \\
~~~
\end{eqnarray}
On the right we have indicated the DDH ``best value" equivalents as a {\it very} rough guide to magnitudes,
with the main points being the strength of the long-range $\Lambda_{1~DDH}^{{}^3S_1-{}^3P_1}$ coupling
(keep in mind that this often contributes only to the exchange channel, which can weaken the contribution
by a factor $\sim 5$ \cite{rev1}) and the weakness of $\Lambda_{1~DDH}^{{}^1S_0-{}^3P_0}$.
Here we have taken $m_\omega \sim m_\rho$, to streamline the expressions:  if one does
not make this simplification, all $\omega$ terms are multiplied by $m_\rho^2/m_\omega^2 \sim 0.97$.

Sometimes the DDH potential is discussed as though it were quite distinct from the nonrelativistic LO potential
of Eq. (\ref{eq:LO}), which can be constructed from symmetries and power-counting in external momenta.
But the expressions above show that this is not the case.  First, in the isoscalar and isotensor channels,
there is a 1-to-1 mapping of DDH weak couplings and LO operator coefficients,
\[ \{\Lambda_{0~DDH}^{{}^1S_0-{}^3P_0}, \Lambda_{0~DDH}^{{}^3S_1-{}^1P_1} \} \leftrightarrow \{ h_\rho^0, h_\omega^0 \}  ~~~~~~~~~~ \{ \Lambda_{2~DDH}^{{}^1S_0-{}^3P_0} \} \leftrightarrow \{ h_\rho^2 \}. \]
In the isovector channel in principle there is an overcompleteless in the LO reduction of the DDH theory
because four weak couplings map onto the two independent operators
\[ \{ \Lambda_{1~DDH}^{{}^1S_0-{}^3P_0} ,\Lambda_{1~DDH}^{{}^3S_1-{}^3P_1} \} \leftrightarrow \{h_\pi^1, h_\rho^1,h_\omega^1,h_\rho^{1\prime}\}. ~~~~~~~~~~~~~\]
However $\Lambda_{1~DDH}^{{}^3S_1-{}^3P_1}$ is completely dominated by $h_\pi^1$, with the contributions
from $h_\rho^1$, $h_\rho^1$ and $h_\rho^{1 \prime}$ contributing at the level
of a few percent.   (In fact customarily $h_\rho^{1\prime} \equiv 0$.)  Consequently, if the DDH potential
were used in processes where the momentum scale $q \rightarrow 0$, one would find that $\Delta I=1$ observables would
depend only on two parameters, $h_\pi^1$ and the combination 
$h_\rho^1 +   h_\omega^1 (g_\omega(2+\chi_S))/(g_\rho(2+\chi_V))$:
with this observation the DDH potential would be functionally equivalent to the most general LO effective theory.
(The momentum scale for most PNC observables is not insignificant, as we will discuss later, so the DDH potential
does take partial account of $P-D$ and other contributions, as discussed in the next section.)

This discussion also further clarifies the motivation for our reanalysis of the Carlson {\it et al.} calculation
of $\vec{p}+p$:  the mapping from effective operator coefficients to DDH parameters involves products
of weak and strong coefficients.  Any consistent global analysis in terms of weak couplings must use a
fixed set of strong coefficients, so the these products are properly determined.

The effective Hamiltonian of Eq.(\ref{eq:LO}) is written in terms of operators of definite isospin, but one can
of course use it to calculate NN observables.  For initial and final states with strong-interaction distorted waves
one finds
\begin{eqnarray}
\langle {}^1S_0| V_{LO}^{PNC} |{}^3P_0 \rangle_{pp} &=& 4\Lambda_{pp}^{{}^1S_0-{}^3P_0} \langle L=0|| {\delta(\vec{r}) \over m_\rho^2} {1 \over i} {\stackrel{\textstyle \rightarrow}{\nabla} \over 2m_N} ||L=1\rangle \nonumber \\
\langle {}^1S_0| V_{LO}^{PNC} |{}^3P_0 \rangle_{nn} &=& 4\Lambda_{nn}^{{}^1S_0-{}^3P_0}\langle L=0|| {\delta(\vec{r}) \over m_\rho^2} {1 \over i} {\stackrel{\textstyle \rightarrow}{\nabla} \over 2m_N} ||L=1\rangle \nonumber \\
\langle {}^1S_0| V_{LO}^{PNC} |{}^3P_0 \rangle_{pn} &=&4\Lambda_{np}^{{}^1S_0-{}^3P_0}\langle L=0|| {\delta(\vec{r}) \over m_\rho^2} {1 \over i} {\stackrel{\textstyle \rightarrow}{\nabla} \over 2m_N} ||L=1\rangle \nonumber \\
\langle {}^3S_1| V_{LO}^{PNC} |{}^1P_1 \rangle_{pn} &=& -{4 \over \sqrt{3}}  \Lambda_{0}^{{}^3S_1-{}^1P_1}\langle L=0|| {\delta(\vec{r}) \over m_\rho^2} {1 \over i} {\stackrel{\textstyle \rightarrow}{\nabla} \over 2m_N} ||L=1\rangle \nonumber \\
\langle {}^3S_1| V_{LO}^{PNC} |{}^3P_1 \rangle_{pn} &=& {4 \over \sqrt{3}} \Lambda_{1}^{{}^3S_1-{}^3P_1}\langle L=0|| {\delta(\vec{r}) \over m_\rho^2} {1 \over i} {\stackrel{\textstyle \rightarrow}{\nabla} \over 2m_N} ||L=1\rangle
\end{eqnarray}
where we have defined
\begin{eqnarray}
\Lambda_{pp}^{{}^1S_0-{}^3P_0} &=& \Lambda_{0}^{{}^1S_0-{}^3P_0} + \Lambda_{1}^{{}^1S_0-{}^3P_0}+{\Lambda_{2}^{{}^1S_0-{}^3P_0} \over \sqrt{6}} \nonumber \\
\Lambda_{nn}^{{}^1S_0-{}^3P_0} &=& \Lambda_{0}^{{}^1S_0-{}^3P_0} - \Lambda_{1}^{{}^1S_0-{}^3P_0}+{\Lambda_{2}^{{}^1S_0-{}^3P_0} \over \sqrt{6}} \nonumber \\
\Lambda_{pn}^{{}^1S_0-{}^3P_0} &=& \Lambda_{0}^{{}^1S_0-{}^3P_0}+\sqrt{{2 \over 3}} ~ \Lambda_{2}^{{}^1S_0-{}^3P_0}
\end{eqnarray}
In terms of quantum number $|(LS)JM_J;TM_T \rangle$, here the $pn$ states have been defined as the normalized
states $|{}^1S_0\rangle_{pn} \equiv |(00)00;10\rangle$, $|{}^1P_1\rangle_{pn} \equiv |(10)1 M_J;00\rangle$, etc.

\section{DDH Potential \& EFTs: A Rosetta Stone}
We have seen, with the customary choice in the DDH potential of $h_\rho^{1 \prime}=0$, the one
redundancy in the potential, when viewed in the S-P limit, is in the isovector channel.  This is the channel
dominated by pion-exchange, and thus the channel where the contact-gradient expansion that comes from
taking the limit $m_\mathrm{meson}\rightarrow  \infty$ will fail first, as a function of increasing momentum transfer.
Following \cite{rev1}, this redundancy
can be expressed as an invariance of the DDH S-P threshold amplitudes under the simultaneous shifts
\begin{eqnarray}
 g_{\pi NN} h_\pi^1 &\rightarrow&g_{\pi NN}  h_{\pi}^1+ \eta \nonumber \\
 g_\rho h_\rho^1 &\rightarrow& g_\rho h_\rho^1 -{\eta~ m_\rho^2 \over \sqrt{2} m_\pi^2}{2 + \mu_S \over 4 + \mu_S+\mu_V} \nonumber \\
 g_\rho h_\omega^1 & \rightarrow&
g_\rho h_\omega^1 +{\eta~ m_\rho^2 \over \sqrt{2} m_\pi^2}{2 + \mu_V \over 4 + \mu_S+\mu_V} 
\end{eqnarray}
This transformation leaves the isosector $^1S_0 \leftrightarrow ^3P_0$ and $^3S_1 \leftrightarrow ^3P_1$ amplitudes invariant.
While the triplet isovector channel contains the pion contribution, the $S-P$ limit provides no information on the range of the interaction:
the pion contribution cannot be disentangled from the $\rho/\omega$ contribution.

The are two $P-D$ $\Delta T=1$ partial waves sensitive
to the pion-range physics
\begin{eqnarray}
^3P_1 - {}^3D_1&:&\Lambda_1^{^3P_1 - ^3D_1} \left\{ -{\overleftarrow{\nabla} \over 2 i  m_N} {\delta(\vec{r}) \over m_\pi^2}
\cdot \left[ (\vec{\sigma}_1 + \vec{\sigma}_2) \otimes \left[ {\overrightarrow{\nabla} \otimes \overrightarrow{\nabla} \over m_\pi^2} \right]_2 \right]_1 \right.\nonumber \\
 &+& \left. \left[ \left[{\overleftarrow{\nabla} \otimes \overleftarrow{\nabla} \over m_\pi^2}\right]_2 \otimes (\vec{\sigma}_1 + \vec{\sigma}_2) \right]_1 \cdot   {\delta(\vec{r}) \over m_\pi^2} {\overrightarrow{\nabla} \over 2 i m_N}  \right\} (\vec{\tau}_1^z-\vec{\tau}_2^z) \nonumber \\
 ^3P_2 -{} ^3D_2&:&\Lambda_1^{^3P_2 - ^3D_2} \left\{ -\left[ {\overleftarrow{\nabla} \over 2 i  m_N} \otimes  (\vec{\sigma}_1 + \vec{\sigma}_2) \right]_2 \cdot {\delta(\vec{r}) \over m_\pi^2}
 \left[ {\overrightarrow{\nabla} \otimes \overrightarrow{\nabla} \over m_\pi^2} \right]_2  \right.\nonumber \\
 &+& \left.  \left[{\overleftarrow{\nabla} \otimes \overleftarrow{\nabla} \over m_\pi^2}\right]_2  \cdot   {\delta(\vec{r}) \over m_\pi^2} \left[ (\vec{\sigma}_1 + \vec{\sigma}_2) \otimes  {\overrightarrow{\nabla} \over 2 i m_N} \right]_2  \right\} (\vec{\tau}_1^z-\vec{\tau}_2^z) \nonumber 
\end{eqnarray}
Here $\otimes$ denotes a tensor product.  These operators come from the next order in the Taylor expansion of the initial and final wave
functions.  The derivation of these terms from a potential model would identify the delta functions in these terms as the $r^2$-weighted moments
of the Yukawa potentials $e^{-m r}/r$, which would consequently emphasize pion-exchange contributions.

In principle, these NLO contributions break the degeneracy among the isovector parameters of the DDH potential.  Given a complete
set of low-energy data that includes some observable with sensitivity to $D$-waves but where momentum scales still
allow a Taylor series approximation, the data will effectively fix the strength of $g_{\pi NN}h_\pi^1$.
The ``sixth degree of freedom" in the DDH potential thus allows one to take account of some average effect of the $^3P_1 - {}^3D_1$ and  $^3P_2 -{} ^3D_2$ channels: 
strong interaction effects that differentiate these channels, such as spin-orbit interactions, would not be
be treated in such a fit.
This view of the DDH potential has much in common so-called hybrid
EFT methods, where conventional potential models are used to generate wave functions, while interactions are expanded
systematically, then evaluated between potential-model wave functions \cite{EFTStar}. In practice, we will argue below that
existing data provide perhaps only two significant constraints on PNC, and thus fall far short of what is needed for even a LO fit,
let alone one that tries to address the finite-range effects of $m_\pi$.

There have been a number of recent treatments of PNC that have developed $S-P$ representations of interactions from a 
``bottom up" approach, developing the $S-P$ amplitudes in the framework of pionless EFT, with the inclusion of one derivative.  
The first such effort was by Zhu et. al \cite{zhu} (though we note EFT approaches have connections to earlier work \cite{dem,dan}).
This treatment retained operators that were related under operator identities analogous to those given in Eq. (\ref{eq:identity}), and
thus is somewhat more difficult to use because of the redundancies.  The long-wavelength form of the resulting PNC Zhu potential is
\begin{eqnarray}
V_{LO}^{Zhu}= -2 {\tilde{{\cal C}}_{6} \over \Lambda_{\chi}^{3}}  i(\vec{\tau}_1\times \vec{\tau}_2)_z (\vec{\sigma}_1+\vec{\sigma}_2)\cdot {1 \over i} \overleftrightarrow{\nabla}_S \delta(\vec{r})+2{{\cal C}_{3} \over \Lambda_{\chi}^3} (\vec{\tau}_1 \cdot \vec{\tau}_2)(\vec{\sigma}_1-\vec{\sigma}_2)\cdot {1 \over i} \overleftrightarrow{\nabla}_A \delta(\vec{r}) \nonumber \\
-2{\tilde{{\cal C}}_{3} \over \Lambda_{\chi}^3} (\vec{\tau}_1\cdot \vec{\tau}_2) i(\vec{\sigma}_1 \times \vec{\sigma}_2) \cdot {1 \over i} \overleftrightarrow{\nabla}_S \delta(\vec{r})+{{\cal C}_{4} \over \Lambda_{\chi}^{3}}
      ({\tau}_1^z+{\tau}_2^z) (\vec{\sigma}_1-\vec{\sigma}_2)\cdot {1 \over i} \overleftrightarrow{\nabla}_A \delta(\vec{r})~~~~~~~~~~ \nonumber \\
-{\tilde{{\cal C}}_{4} \over \Lambda_{\chi}^{3}} ({\tau}_1^z+{\tau}_2^z) i (\vec{\sigma}_1 \times \vec{\sigma}_2)\cdot {1 \over i} \overleftrightarrow{\nabla}_S \delta(\vec{r}) + 2 \sqrt{6} {{\cal C}_{5} \over \Lambda_{\chi}^{3}} (\tau_1 \otimes \tau_2)_{20} (\vec{\sigma}_1-\vec{\sigma}_2)\cdot {1 \over i} \overleftrightarrow{\nabla}_A \delta(\vec{r})~~ \nonumber \\
-2 \sqrt{6} {\tilde{{\cal C}}_{5} \over \Lambda_{\chi}^{3}} (\tau_1 \otimes \tau_2)_{20} i (\vec{\sigma}_1 \times \vec{\sigma}_2) \cdot {1 \over i} \overleftrightarrow{\nabla}_S \delta(\vec{r}) +2 {{\cal C}_{1} \over \Lambda_{\chi}^{3}} (\vec{\sigma}_1-\vec{\sigma}_2)\cdot {1 \over i} \overleftrightarrow{\nabla}_A \delta(\vec{r})~~~~~~~~~~~~ \nonumber \\
-2{ \tilde{{\cal C}}_{1} \over \Lambda_{\chi}^{3}} i (\vec{\sigma}_1 \times \vec{\sigma}_2) \cdot {1 \over i} \overleftrightarrow{\nabla}_S \delta(\vec{r})
+ {{\cal C}_{2} \over \Lambda_{\chi}^{3}} ({\tau}_1^z+{\tau}_2^z) (\vec{\sigma}_1-\vec{\sigma}_2)\cdot {1 \over i} \overleftrightarrow{\nabla}_A \delta(\vec{r})~~~~~~~~~~~~~~~~~~~ \nonumber \\
-{\tilde{{\cal C}}_{2} \over \Lambda_{\chi}^{3}} ({\tau}_1^z+{\tau}_2^z) i (\vec{\sigma}_1 \times\vec{\sigma}_2) \cdot {1 \over i} \overleftrightarrow{\nabla}_S \delta(\vec{r}) + {({\cal C}_{2}-{\cal C}_{4}) \over \Lambda_{\chi}^{3}} ({\tau}_1^z-{\tau}_2^z) (\vec{\sigma}_1+\vec{\sigma}_2) \cdot {1 \over i} \overleftrightarrow{\nabla}_A \delta(\vec{r}) \nonumber \\
\end{eqnarray}

Girlanda\cite{gir} then provided a treatment that addressed the necessary operator identities,
generating a potential with the requisite 
 five low energy constants.  The analysis begins with the twelve operators
\begin{eqnarray}
{\cal
O}_1&=&\bar{\psi}\gamma^\mu\psi\bar{\psi}\gamma_\mu\gamma_5\psi\nonumber\\
\tilde{{\cal
O}}_1&=&\bar{\psi}\gamma^\mu\gamma_5\psi\partial^\nu(\bar{\psi}\sigma_{\mu\nu}\psi)\nonumber\\
{\cal
O}_2&=&\bar{\psi}\gamma^\mu\psi\bar{\psi}\tau_3\gamma_\mu\gamma_5\psi\nonumber\\
\tilde{{\cal
O}}_2&=&\bar{\psi}\gamma^\mu\gamma_5\psi\partial^\nu(\bar{\psi}\tau_3\sigma_{\mu\nu}\psi)\nonumber\\
{\cal
O}_3&=&\bar{\psi}\tau_a\gamma^\mu\psi\bar{\psi}\tau^a\gamma_\mu\gamma_5\psi\nonumber\\
\tilde{{\cal
O}}_3&=&\bar{\psi}\tau_a\gamma^\mu\gamma_5\psi\partial^\nu(\bar{\psi}\tau^a\sigma_{\mu\nu}\psi)\nonumber\\
{\cal
O}_4&=&\bar{\psi}\tau_3\gamma^\mu\psi\bar{\psi}\gamma_\mu\gamma_5\psi\nonumber\\
\tilde{{\cal
O}}_4&=&\bar{\psi}\tau_3\gamma^\mu\gamma_5\psi\partial^\nu(\bar{\psi}\sigma_{\mu\nu}\psi)\nonumber\\
{\cal O}_5&=&{\cal
I}_{ab}\bar{\psi}\tau_a\gamma^\mu\psi\bar{\psi}\tau_b\gamma_\mu\gamma_5\psi\nonumber\\
\tilde{{\cal
O}}_5&=&{\cal I}_{ab}\bar{\psi}\tau_a\gamma^\mu\gamma_5\psi\partial^\nu(\bar{\psi}\tau_b\sigma_{\mu\nu}\psi)\nonumber\\
{\cal
O}_6&=&i\epsilon_{ab3}\bar{\psi}\tau_a\gamma^\mu\psi\bar{\psi}\tau_b\gamma_\mu\gamma_5\psi\nonumber\\
\tilde{{\cal
O}}_6&=&i\epsilon_{ab3}\bar{\psi}\tau_a\gamma^\mu\gamma_5\psi\partial^\nu(\bar{\psi}\tau_b\sigma_{\mu\nu}\psi)
\end{eqnarray}
generating the general PV NN Lagrangian
\begin{equation}
{\cal L}_{PVNN}=\sum_{j=1}^{6}[{\cal G}_i{\cal O}_i+\tilde{\cal G}_i{\cal O}_i].
\end{equation}
Then, with the use of Fierz transformations and the free-particle
equation of motion, six conditions relating these operators were identified
\begin{eqnarray}
\label{eq:gl}
{\cal O}_3&=&{\cal O}_1\nonumber\\
{\cal O}_2-{\cal O}_4&=&2{\cal O}_6\nonumber\\
\tilde{{\cal O}}_3+3\tilde{{\cal O}}_1&=&2m_N({\cal O}_1+{\cal
O}_3)\nonumber\\
\tilde{{\cal O}}_2+\tilde{{\cal O}}_4&=&m_N({\cal O}_2+{\cal
O}_4)\nonumber\\
\tilde{{\cal O}}_2-\tilde{{\cal O}}_4&=&-2m_N{\cal O}_6-\tilde{{\cal
O}}_6\nonumber\\
\tilde{{\cal O}}_5&=&{\cal O}_5
\end{eqnarray}
Finally, using the feature that the operators ${\cal O}_6$ and
$\tilde{{\cal O}}_6$ have the {\it same} form in the lowest order
nonrelativisitic expansion, one determines an effective (pionless) Lagrangian, which reduces to the nonrelativistic form
\begin{eqnarray} 
V_{LO}^{Girlanda} = \left[ -2 \tilde{\cal G}_1 \right]~ {1 \over i} \overleftrightarrow{\nabla}_S~ \delta(\vec{r}) \cdot i (\vec{\sigma}_1 \times \vec{\sigma}_2) + \left[ 2 {\cal G}_1 \right]~ {1 \over i} \overleftrightarrow{\nabla}_A~ \delta(\vec{r}) \cdot (\vec{\sigma}_1-\vec{\sigma}_2) \nonumber \\
  + \left[ {\cal G}_2 \right]~ {1 \over i} \overleftrightarrow{\nabla}_A~\delta(\vec{r}) \cdot (\vec{\sigma}_1-\vec{\sigma}_2) (\tau_1^z+\tau_2^z) + \left[2 {\cal G}_6 \right]~ {1 \over i} \overleftrightarrow{\nabla}_A~\delta(\vec{r}) \cdot (\vec{\sigma}_1+\vec{\sigma}_2) (\tau_1^z- \tau_2^z) \nonumber \\
  + \left[-2 \sqrt{6}{\cal G}_5 \right]~ {1 \over i} \overleftrightarrow{\nabla}_A~\delta(\vec{r}) \cdot (\vec{\sigma}_1-\vec{\sigma}_2) (\tau_1\otimes \tau_2)_{20} ~~~~~~~~~~~~~~~~~~~~~
   \end{eqnarray}
(As was done by Phillips, Schindler, and Springer\cite{phi}, in Eq. (\ref{eq:gl}) the factor of $1/\Lambda_{\chi}^{3}$ used by Girlanda has been absorbed into the coefficients, making them dimensional.)  

\begin{table}[h]
\caption{The coefficients of the S-P PNC potential of Eq. (\ref{eq:LO}) in the DDH potential, Girlanda, and Zhu descriptions.
Note that multiplicative factors of  $2m_N m_\rho^2$  and $2m_N m_\rho^2/\Lambda_\chi^3$ must be applied to the
Girlanda and Zhu entries, respectively, to obtain the dimensionless coefficients $\Lambda$, e.g.,
$\Lambda_{0~DDH}^{{}^1S_0-{}^3P_0}=2({\cal G}_1 +\tilde{{\cal G}}_1)[2 m_N m_\rho^2] = 
2({\cal C}_1+\tilde{{\cal C}}_1+ {\cal C}_3+\tilde{{\cal C}}_3)[2 m_N m_\rho^2/\Lambda_\chi^3]$.}
\label{tab:ros}
\begin{center}
\begin{tabular}{cccc}
\hline
& & & \\[-.3cm]
Coeff & DDH & Girlanda & Zhu  \\[.2cm]
\hline
 & & & \\[-.2cm]
$\Lambda_{0~DDH}^{{}^1S_0-{}^3P_0}$ &$ - g_\rho h_\rho^0 (2$+$ \chi_V) -g_\omega h_\omega^0(2$+$\chi_S) $ & $2({\cal G}_1 $+$\tilde{{\cal G}}_1)$ & $2({\cal C}_1$+$\tilde{{\cal C}}_1$+$ {\cal C}_3$+$\tilde{{\cal C}}_3) $\\ [.2cm]
$\Lambda_{0~DDH}^{{}^3S_1-{}^1P_1} $&  $g_\omega h_\omega^0 \chi_S- 3g_\rho h_\rho^0 \chi_V $ &$ 2({\cal G}_1$-$\tilde{{\cal G}}_1) $& $2({\cal C}_1$-$\tilde{{\cal C}}_1$-$3 {\cal C}_3$+$3\tilde{{\cal C}}_3)$ \\[.2cm]
$\Lambda_{1~DDH}^{{}^1S_0-{}^3P_0}$ & $-g_\rho h_\rho^1 (2$+$ \chi_V)- g_\omega h_\omega^1 (2$+$\chi_S)$  & $ {\cal G}_2$ & $({\cal C}_2$+$\tilde{{\cal C}}_2$+$ {\cal C}_4$+$\tilde{{\cal C}}_4)$\\[.2cm]
$\Lambda_{1~DDH}^{{}^3S_1-{}^3P_1} $&  $  {1 \over \sqrt{2}}  g_{\pi NN} h_\pi^1 \left( {m_\rho \over m_\pi} \right)^2$+$ g_\rho (h_\rho^1$-$h_\rho^{1\prime})-g_\omega h_\omega^1$  & $2{\cal G}_6 $& $(2\tilde{{\cal C}}_6$+$ {\cal C}_2$-${\cal C}_4))$  \\[.2cm]
$\Lambda_{2~DDH}^{{}^1S_0-{}^3P_0} $& $ -g_\rho h_\rho^2 (2$+$ \chi_V)$ & $-2 \sqrt{6}{\cal G}_5 $& $2 \sqrt{6}({\cal C}_5$+$\tilde{{\cal C}}_5)$ \\[.2cm]
\hline
\end{tabular}
\end{center}
\end{table}

Returning to the ``canonical form" of the $S-P$ contact potential in terms of the partial-wave operators of Eq. (\ref{eq:LO}), the relationships between
the DDH, Girlanda, and Zhu forms of that potential can be summarized in terms of coefficients of that potential, as shown in Table \ref{tab:ros}.  In using this table it should be remembered that the DDH results include the assumption 
that a one-boson exchange potential operates between strongly interacting initial and final nuclear states.  There are
contributions from crossed-pion diagrams and delta intermediate states that cannot be factored in this way.
Inclusion of such terms would alter the mapping between coefficients and meson couplings shown in
the table.  See, for example, Refs. \cite{simonius,zhu}.

\section{Observables and Momentum Scales}
The decomposition of the PNC interaction into SP amplitudes provides an interesting way to think
about the need for additional experimental constraints on the PNC interaction.   This type of
approach -- envisioning experiments that might constrain the  five degrees of freedom of an $S-P$ potential, valid near threshold --
dates back to the work on Danilov \cite{dan}.  (See also Ref. \cite{dem}.)

One can approach the problem of obtaining five experimental constraints on the PNC potential 
with varying degrees of realism.  For example, Danilov suggested a treatment where the strong-interaction
input would be limited to the $^3S_1$ and $^1S_0$ strong phase shifts.  The parameters derived from fitting
experiment in this limit would encode and thus entangle weak and strong physics: for example, the
strong short-range repulsion that carves out the $r \sim 0.5$ fm hole in the nucleon-nucleon correlation
function would be entangled with the short-range weak physics the DDH potential attributes to
$\rho$ and $\omega$ exchange.  Yet the effective couplings derived from fitting data would still
provide a valid parameterization of low-energy weak NN interactions, and once those parameters
were determined, could be used to make predictions.

Several such ``unitarized" strong phase-shift methods designed to satisfy the generalized 
Watson theorem were developed and employed in calculations of PNC observables
such as $A_L(\vec{p}+p)$ in the 1970s and 80s \cite{henley1975,oka1981}.  The
inadequacies of such approaches in applications to low-energy observables such as $A_L(\vec{p}+p)$
were noted in the 1980s \cite{rev1}.  The importance of treating the strong distortion of the
partial waves was first demonstrated in explicit calculations by Driscoll and Miller \cite{dm}.
The need for distorted waves follows from simple considerations:  our highest precision 
measurements of $A_L(\vec{p}+p)$ were performed at 13.6 and 45 MeV, or at center-of-mass
relative momenta of 80 and 145 MeV.  A treatment like that envisioned by Danilov in which
the information about the strong interaction is encoded entirely in the scattering length cannot
be successfully applied to these data:  the range of validity is determined not by the naive estimate
of $m_\pi$, but by the anomalously large scattering lengths of the $nn,~np,$ and $pp$ channels of
$\sim$ 20 f.  Using the effective range expansion, one find that the neglect of the range,
or $r_0$,  in
\begin{equation}
q \cot{\delta(q)} = -{1 \over a_0} + {r_0 \over 2} q^2 + \dots
\end{equation}
produces 100\% errors at $q \sim 34$ MeV/c.  This constraint would require elementary NN experiments to be performed at scattering
energies $\lesssim 2$ MeV, which appears to be impractical except, potentially, in future spin rotation
experiments in parahydrogen.

While measuring NN scattering observables at very low energies may be prohibitively
difficult, one might propose alternative measurements of near-threshold PNC observables in few-nucleon
system, such as neutron spin rotation in $^4$He.  Unfortunately the effective center-of-mass momentum
relevant to NN partial-wave analyses is generally not determined by the external kinematics in such
reactions, but rather by the nuclear Fermi momentum, typically $\sim$ 200 MeV$ > m_\pi$.
This is the characteristic scale of the momentum flow in the exchange term, which
often dominates the PNC interaction, as Barton's theorem restricts direct terms.  There is a nice
pedagogical treatment of these effects in \cite{rev1} (see Sec. 6.1), where a mean-field approximation
to $V_{12}^{PNC}$ is derived in a Fermi gas model, showing analytically that exchange terms make important
contributions to every $S-P$ amplitude, are the only source of sensitivity to $h_\pi^1$, and tend
to dominate couplings like $h_\rho^0$ in light isoscalar nuclei.

Consequently, we are stuck with the need for evaluating PNC observables in a framework that
treats strong interaction distortions.  This framework is not provided by Danilov or pionless EFT
approaches in which strong interaction input is limited to scattering lengths.  Pionful EFTs 
could be contemplated, but their development for strongly
interacting $A>2$ systems has proven to be challenging due to the anomalous scales mentioned above,
apparent from nuclear binding energies:
this was the origin of Weinberg's proposal to distinguish irreducible graphs
that involve no infrared enhancements, treating these by chiral perturbation theory, from reducible graphs,
which involve the strong potential $V$ in combination with infrared-enhanced propagators,
requiring all-order summations.  Such considerations have led to hybrid EFT strategies where
wave functions are generated from potentials that have been tuned to the anomalous nuclear scales, 
but where the interactions evaluated between such wave functions are developed through EFT.
This appears to us to be a viable strategy for modern treatments of PNC, with the coefficients of
the five $S-P$ operators providing a model-independent description of the PNC NN interaction.
Such a theoretical program would be timely if new data soon become available.

This leads us to the question of what has been measured vs. what might be measured (or calculated)
in the near future to further our understanding of hadronic PNC.  First we consider the constraints illustrated in Fig. 3,
where we take all calculations from \cite{rev1} except as noted:

\begin{enumerate}[leftmargin=0cm,itemindent=.5cm]
\item  The longitudinal $\vec{p}+p$ asymmetry at 13.6 \cite{ghs}, 15 \cite{nagle}, 45 \cite{ght},  221 \cite{TRIUMF} MeV: \\

\noindent
The new combined analysis described in Sec. 4 yields the $S-P$ constraint
\[ \Lambda_{pp}^{^1S_0-^3P_0} \equiv \Lambda_{0}^{^1S_0-^3P_0} +\Lambda_{1}^{^1S_0-^3P_0} +{\Lambda_{2}^{^1S_0-^3P_0}  \over \sqrt{6}}= ( 4.19 \pm 0.43 ) \times 10^{-5}~(68\% \mathrm{c.l.}) \]
It also yields a weaker constraint on the $^3P_2-^1D_2$ amplitude which we  give here in terms of DDH couplings
\[ g_\rho h_\rho^{pp} \chi_V +g_\omega h_\omega^{pp} \chi_S = -(4.4 \pm 1.6) \times 10^{-5}~(68\% \mathrm{c.l.}) \]
\item The longitudinal $\vec{p}+^4$He asymmetry at 46 MeV\cite{ghu}
\begin{eqnarray}
A_L(\vec{{\rm p}}\alpha,\,46\,{\rm
MeV})&=&- 0.025 g_{\pi NN} h_\pi^1 +0.050 g_\rho h_\rho^0+0.017 g_\rho h_\rho^1 +0.007 g_\omega h_\omega^0  \nonumber \\
&& +0.007 g_\omega h_\omega^1 \nonumber \\
&\sim &-0.00355 \Lambda_0^{^1S_0-^3P_0}
 -0.00317 \Lambda_1^{^1S_0-^3P_0} \nonumber \\
 && -0.00268\Lambda_0^{^3S_1-^1P_1}-0.00114 \Lambda_1^{^3S_1-^3P_1} \nonumber \\
 &=&  -(3.3\pm 0.9)\times 10^{-7}
\end{eqnarray}
\item The circular polarization of the $\gamma$-rays omitted in the decay of excited $^{18}$F.  The data from the various experiments discussed earlier can be
combined to determine $P_\gamma(^{18}F) < (1.2 \pm 3.9) \times 10^{-4}$.  As the relationship of this limit to the underlying weak parameters
depends on a mixing ratio of known magnitude but unknown sign \cite{rev1}, one finds in terms of DDH couplings
\[ |P_\gamma(^{18}\mathrm{F})| = |326 g_{\pi NN}h_\pi^1 -176 g_\rho h_\rho^1 -100 g_\omega h_\omega^1|   <  5.1 \times 10^{-4} \]
where the nuclear matrix element has been taken from axial-charge $\beta$-decay measurements.  This result can be recast as an
approximate constraint on $S-P$ coefficients
\[  |P_\gamma(^{18}\mathrm{F})| \sim 15.0 |\Lambda_1^{^3S_1-^3P_1}+2.42 \Lambda_1^{1S_0-^3P_0} | <  5.1 \times 10^{-4}. \]
As discussed previously, there is also a significant result from lattice QCD that provides an important
comparison for the $^{18}$F result \cite{wasem}
\[ h_\pi^1(\mathrm{LQCD~connected}) = \left(1.10 \pm 0.51 \mathrm{(stat)} ^{+0.51}_{-0.60} \mathrm{(sys)}\right) \times 10^{-7} \]
\item The $\gamma$-ray asymmetry in $^{19}$F
can be converted to the following constraint on weak couplings using nuclear matrix elements calibrated
by axial-charge $\beta$ decay (see earlier discussions on associated uncertainties),
\begin{eqnarray}
A_\gamma(^{19}\mathrm{F}) &=&-7.00 g_{\pi NN} h_\pi^1+12.2 g_\rho h_\rho^0+3.65 g_\rho h_\rho^1 +2.31 g_\omega h_\omega^0+2.02 g_\omega h_\omega^1\nonumber \\
&\sim& -1.12 \Lambda_0^{^1S_0-^3P_0}- 0.75 \Lambda_1^{^1S_0-^3P_0}-0.48 \Lambda_0^{^3S_1-^1P_1} -0.32 \Lambda_1^{^3S_1-^3P_1}\nonumber \\
&=& -(7.4 \pm 1.9) \times 10^{-5} 
 \end{eqnarray}
 These constrains are similar to those for $\vec{p}+^4$He, another odd-proton system.
 \end{enumerate}
 
 As we have discussed previously, the results above, which are shown graphically in Fig. 3,  are those 
 that we feel can be reliably interpreted.
 There is a great need for additional measurements, particularly in NN and few-body
 systems where the associated strong-interaction effects can be handled well:
 \begin{enumerate}[leftmargin=0cm,itemindent=.5cm]
 \item The photon asymmetry in $\vec{{ n}}{ p}\rightarrow d\gamma$\cite{cavaignac,bow}
\begin{eqnarray}
A_\gamma(\vec{n}p\rightarrow d + \gamma)&=& -0.0080 g_{\pi NN} h_\pi^1 -0.0005 g_\rho h_\rho^1 +0.0005 g_\omega h_\omega^1 \nonumber \\
&\sim& -(3.7 \times 10^{-4}) \Lambda_1^{^3S_1-^3P_1}\nonumber \\
&=&\left\{ \begin{array}{c}(0.6\pm 2.1) \times 10^{-7}  \\
(-1.2\pm 1.9\pm 0.2) \times 10^{-7} \end{array} \right. 
\end{eqnarray}
The SNS continuation of the program begun in \cite{bow} may be able to reach a sensitivity of $\sim 10^{-8}$.  The lattice QCD prediction of $A_\gamma \sim -1.2 \times 10^{-8}$ and the $^{18}$F results suggest that such sensitivity will be necessary to see a nonzero signal.
\item  The neutron spin rotation in $^4$He\cite{sno}.  Using \cite{flambaum}, we find
\begin{eqnarray}
{d\phi^{{n}\alpha}\over dz}&=& \left[-0.072 g_{\pi NN} h_\pi^1-0.115 g_\rho h_\rho^0+0.039 g_\rho h_\rho^1-0.026 g_\omega h_\omega^0+ \right. \nonumber \\
&&\left.~~~~~ +0.026 g_\omega h_\omega^1\right] {\rm rad/m}  \nonumber \\
&\sim& \left[ 0.0138 \Lambda_0^{^1S_0-^3P_0}-0.0087 \Lambda_1^{^1S_0-^3P_0} + 0.0033 \Lambda_0^{^3S_1-^1P_1}\right. \nonumber \\
&&\left.~~~~~ -0.0033 \Lambda_1^{^3S_1-^3P_1} \right] {\rm rad/m} \nonumber \\
&=& (1.7\pm 9.1\pm 1.4)\times 10^{-7} {\rm rad/m}
\end{eqnarray}
This observable is an isospin complement of  $A_\gamma(^{19}$F) and $A_L(\vec{p}\alpha)$, so it is unfortunate
that the sensitivity necessary to obtain an important, orthogonal constraint has so far not been reached.
\item The circular polarization of the photons emitted in the capture of unpolarized thermal neutrons
by protons \cite{lob1,lob2}
\begin{eqnarray}
P_\gamma(np\rightarrow d+ \gamma)
&=&  -0.011 g_\rho h_\rho^0-0.0088 g_\rho h_\rho^2 +0.0001 g_\omega h_\omega^0 \nonumber \\
  &\sim& -0.00012 \Lambda_0^{^1S_0-^3P_0}+0.00105 \Lambda_0^{^3S_1-^1P_1}+0.00154 \Lambda_2^{^1S_0-^3P_0} \nonumber \\
  &=& (1.8 \pm 1.8) \times 10^{-7} 
\end{eqnarray} 
\item The analyzing power for $\vec{p}+d$.  The existing limit was obtained at 15 MeV, while the theoretical estimate comes from \cite{dbg}:
\begin{eqnarray}
A_L(\vec{p}+d)\Big|_{15 \mathrm{~MeV}} &=&-0.0171 g_{\pi NN} h_\pi^1+0.0085 g_\rho h_\rho^0+0.0035 g_\rho h_\rho^1 \nonumber \\
&&+0.002 g_\omega h_\omega^0+0.0015 g_\omega h_\omega^1  \nonumber \\
&\sim&-0.0010 \Lambda_0^{^1S_0-^3P_0}-.0007 \Lambda_1^{^1S_0-^3P_0} \nonumber \\
&&-0.0002 \Lambda_0^{^3S_1-^1P_1}-0.0008 \Lambda_1^{^3S_1-^3P_1} \nonumber \\
&=&-(0.35\pm 0.85) \times 10^{-7}
\end{eqnarray}
\item The gamma-ray asymmetry 
\begin{eqnarray}
A_\gamma(\vec{n}+d \rightarrow t+ \gamma) &=&0.051 g_{\pi NN} h_\pi^1 -0.12 g_\rho h_\rho^0 +0.036 g_\rho hrho^1 +0.020 g_\rho h_\rho^2 \nonumber \\
&&-0.027 g_\omega h_\omega^0 +0.007 g_\omega h_\omega^1 \nonumber \\
&\sim& 0.0139 \Lambda_0^{^1S_0-^3P_0} -0.0055 \Lambda_1^{^1S_0-^3P_0} +0.0037 \Lambda_0^{^3S_1-^1P_1}\nonumber  \\
&&+ 0.0024 \Lambda_1^{^3S_1-^3P_1}-0.0035 \Lambda_2^{^1S_0-^3P_0} 
\end{eqnarray}
A measurement of this asymmetry was reported some years ago, but is widely disregarded because of its size \cite{avenier}.
\item The as yet unmeasured neutron spin rotation in hydrogen
\begin{eqnarray}
{d\phi^{{n} \mathrm{H}}\over dz}&=&\left[ -0.23 g_{\pi NN} h_\pi^1-0.082 g_\rho h_\rho^0-0.011 g_\rho h_\rho^1-0.090 g_\rho h_\rho^2 \right. \nonumber \\
&&\left.  -0.027 g_\omega h_\omega^0 +0.011 g_\omega h_\omega^1\right] \mathrm{rad/m} \nonumber \\
&=&\left[  0.015 \Lambda_0^{^1S_0-^3P_0} -0.011\Lambda_1^{^3S_1-^3P_1}+0.016 \Lambda_2^{^1S_0-^3P_0}\right] \mathrm{rad/m}
\end{eqnarray}
\end{enumerate}
Other few-nucleon observables have been considered in the literature (though we are aware of no measurements).  Ideas include
the longitudinal asymmetry for $^3$He$(\vec{n},p)^3$H \cite{viviani};
the photon asymmetry where unpolarized neutrons are scattered off a polarized deuterium target, $A_\gamma(n\vec{d})$; neutron spin rotation in a deuterium target \cite{hg}; and the
capture of circularly polarized photons on deuterium $\vec{\gamma} d \rightarrow np$.

\section{Outlook and Summary}
After more than two decades during which few new results became available, we are beginning once again to make progress in understanding how the
$\Delta S=0$ weak interaction operates among strongly interacting nucleons.  First we argued here that if the existing experimental results
yielding nonzero values for PNC observables are
treated in a consistent formalism, the agreement among them is really quite good:  The $\vec{p}+p$, $^{18}$F, $^{19}$F, and $\vec{p}+^4$He 
results combine to suggest a ratio of isoscalar-to-isovector strengths about a factor of six larger than the ``best value" benchmark
of DDH.  The tension that had existed among the results appears to have arisen from comparisons that did not utilize a common
set of strong meson-nucleon vertices.

Second, we have the first tentative confirmation of the most puzzling result in the field, the indication from $^{18}$F that $h_\pi^1$
is at least a factor of three below the nominal DDH ``best value."   The evidence supporting this result has come from theory,
the first lattice QCD calculation of a weak meson-nucleon coupling.  The lattice QCD value for $h_\pi^1$ is consistent
with the $^{18}$F upper bound.  While the result is tentative -- calculations at the physical pion mass with the inclusion of disconnected
contributions remain to be done -- large changes are not expected, based on current estimates of the calculation's statistical
and systematic uncertainties.

Third, as lattice QCD gives an $h_\pi^1$ near the upper bound of the $^{18}$F band, the
ongoing SNS experiment on $\vec{n}+p \rightarrow d + \gamma$ may succeed in measuring isovector PNC, if it reaches its
precision goal of $\sim 10^{-8}$.  This would give us a direct experimental value for $h_\pi^1$ or, equivalently,
$\Lambda_1^{^3S_1-^3P_1}$.

Fourth, there is a very good prospect that $h_\rho^2$ could be calculated rather precisely from lattice QCD: this contribution has
only connected pieces.  This would determine $\Lambda_2^{^1S_0-^3P_0}$ and, consequently, allow one to extract from $\vec{p}+p$
much sharper constraints on $\Lambda_0^{^1S_0-^3P_0}+ \Lambda_1^{^1S_0-^3P_0}$.  Our calculation combining
 the 13.6, 45, and 221 MeV $\vec{p}+p$ results could be redone, without the loss of sensitivity
that comes from marginalizing over possible values of the unknown parameter  $\Lambda_2^{^1S_0-^3P_0}$.

This review has also tried to emphasize the close relationship between the various formulations of hadronic PNC, 
whether based on potentials such as DDH \cite{ddh} or pionless effective field theory,
such as the calculations of Zhu {\it et al.} \cite{zhu}, Girlanda \cite{gir}, and Phillips {\it et al.} \cite{phi}.  The common language
is the set of five S-P operator coefficients that provide a model-independent parameterization of the most general low-energy PNC interaction.
Operationally, the DDH potential differs from pionless EFT interactions only through the inclusion of a sixth parameter that can mock up the effects
of long-range pion-exchange in inducing $^3P-{}^3D$ transitions; alternatively, that degree of freedom can be removed through
a constraint, to obtain an analog of pionless EFT, as was done in \cite{rev1}.

With the exception of the $\vec{p}+p$ result at 221 MeV, the data we have utilized in this review can be represented well in
calculations that employ the $S-P$ operators.  However, we have also stressed the importance of embedding that operator
between realistic strong-interaction wave functions:  momentum transfers in many of the processes of interest are characteristric
of the Fermi momentum, and are thus well beyond the range of validity of Danilov or EFT treatments that limit strong interaction
input to the singlet and triplet scattering lengths.   The importance of distorted waves
for accurate calculations of low energy $\vec{p}+p$, for example, has been known for many years.   The development of
pionful theories to systems with $A>2$ has proven a great challenge, reflecting the anomalously low scale of nuclear
binding energies and the associated difficulties with infrared enhancements.   The kind of approach we envision
succeeding in PNC studies combines a model-independent description of the weak interaction, in the manner of pionless EFT,
with state-of-the-art wave functions taken from modern potentials.  This is sometimes termed hybrid EFT.

Finally, with lattice QCD now making contributions and with a new experimental program underway at the SNS,
 it may be time for a more coordinated theory effort on PNC.  In the previous 
section we have described the dependence of PNC observables on the underlying
$S-P$ operator coefficients.  Many of the calculations used in that section, however, are dated and should be revisited.
One purpose of this review is to set the stage for such a comprehensive effort.   The formalism used here
emphasizes the common features of existing PNC treatments, whether based on potentials or constructed
as pionless EFTs.   We believe a treatment that exploits the $S-P$ operator coefficients as a ``Rosetta stone" for PNC,
but evaluates operators between wave functions that are constructed from the best modern potentials, applied systematically
to all of the relevant few-nucleon systems,would address
the needs of the experimental community.  The simplicity of the $S-P$ operator formalism would be retained, but the
realism achievable with modern potential treatments of the strong interaction would not be sacrificed.\\

\begin{center}
{\bf\Large Acknowledgements}
\end{center}
The work of WCH is supported in part by the US Department of Energy under DE-SC00046548 at Berkeley and 
DE-AC02-98CH10886 at LBL, and that of BRH is supported in part by the National Science Foundation under  PHY-0855119.


\begin{thebibliography}{99}
\bibitem{ley} T.D. Lee and C.N. Yang, Phys. Rev. {\bf 104}, 822 (1956).
\bibitem{amb} C.S. Wu et al., Phys. Rev. {\bf 105}, 1413 (1957).
\bibitem{nta} N. Tanner, Phys. Rev. {\bf 107}, 1233 (1957).
\bibitem{hfe} K.S. Krane et al., Phys. Rev. Lett. {\bf 26}, 1579 (1971); Phys. Rev. {\bf C4}, 1906 (1971).
\bibitem{lae} V.W. Yuan et al., Phys. Rev. {\bf C44}, 2187 (1991); V.P. Alfimenko et al., Nucl. Phys. {\bf A398}, 93 (1983); Y. Masuda et al., Nucl. Phys. {\bf A504}, 269 (1989).
\bibitem{rev1} E.G. Adelberger and W.C. Haxton, Ann. Rev. Nucl. Part. Sci. {\bf 35}, 501 (1985).
\bibitem{rev2} W. Haeberli and B.R. Holstein, in {\it Symmetries and Fundamental Interactions in Nuclear Physics}, ed. E. Henley and W. Haxton, World Scientific, Singapore (1995), p. 17-66.
\bibitem{rev3} M.J. Ramsey-Musolf and S.A. Page, Ann. Rev. Nucl. Part. Sci. {\bf 56},1 (2006).
\bibitem{bd} J. D. Bjorken and S. D. Drell, {\it Relativistic Quantum Mechanics}, McGraw-Hill, New York (1964).
\bibitem{ghs} P.D. Evershiem et al., Phys. Lett. {\bf B256}, 11 (1991).
\bibitem{nagle} D.E. Nagle et al., AIP Conf. Proc. 51 (AIP, New York, 1978), p 224.
\bibitem{ght} R. Balzer et al., Phys. Rev. Lett. {\bf 44}, 699 (1980) and Phys. Rev. {\bf C30}, 1409 (1984);
S. Kistryn et al., Phys. Rev. Lett. {\bf 58}, 1616 (1987).
\bibitem{TRIUMF} A.R. Berdoz et al., Phys. Rev. Lett. {\bf 87}, 272301 (2001) and Phys. Rev. {\bf C68} (2003) 034004.
\bibitem{cavaignac} J. F. Cavaignac, B. Vignon, and R. Wilson, Phys. Lett. {\bf B67}, 148 (1977).
\bibitem{bow} M.T. Gericke et al., Phys. Rev. {\bf C83}, 015505 (2011).
\bibitem{lob1} V.M. Lobashov et al. Nucl. Phys. {\bf A197}, 241 (1972).
\bibitem{lob2} V.A. Knyaz'kov et al., Nucl Phys. {\bf A417}, 209 (1984).
\bibitem{viviani} M. Viviani et al., Phys, Rev. {\bf C82}, 044001 (2010).
\bibitem{nollett} K. M. Nollet et al., Phys. Rev. Lett. {\bf 99}, 022502 (2007).
\bibitem{ghu} J. Lang et al., Phys. Rev. Lett. {\bf 54}, 170 (1985); R. Henneck et al., Phys. Rev. Lett. {\bf 48}, 725 (1982).
\bibitem{sno} W.M. Snow et al., Phys. Rev. {\bf C83} 022501 (2011).
\bibitem{nagle2} D.E. Nagle et al., AIP Conf. Proc. 51 (AIP, New York, 1979) p. 24.
\bibitem{ghk} C.A. Barnes et al., Phys. Rev. Lett. {\bf 40}, 840 (1978).
\bibitem{ghl} M. Bini et al., Phys. Rev. Lett. {\bf 55}, 795 (1985).
\bibitem{ghm} G. Ahrens et al., Nucl. Phys. {\bf A390}, 496 (1982).
\bibitem{ghn} S.A. Page et al., Phys. Rev. {\bf C35}, 1119 (1987).
\bibitem{hax81} W.C. Haxton, Phys. Rev. Lett. {\bf46}, 698 (1981).
\bibitem{gho} E.G. Adelberger et al., Phys. Rev. {\bf C27}, 2833 (1983).
\bibitem{ghp} K. Elsener et al., Nucl. Phys. {\bf A461}, 579 (1987); Phys. Rev. Lett. {\bf 52}, 1476 (1984).
\bibitem{ghq} K.A. Snover et al., Phys. Rev. Lett. {\bf 41}, 145 (1978).
\bibitem{ghr} E.D. Earle et al., Nucl. Phys. {\bf A396}, 221 (1983).
\bibitem{zeld} Ya. B. Zeldovich, Sov. Phys. JETP {\bf 6}, 1184 (1958) and citations therein.
\bibitem{khrip} I. B. Khriplovich, {\it Parity Nonconservation in Atomic Phenomena}, Gordon and Breach, Philadelphia (1991).
\bibitem{ghv} P. Vetter et al., Phys. Lett. {\bf B74}, 2658 (1995).
\bibitem{ghw} N.H. Edwards et al. Phys. Rev. Lett. {\bf 74} 2654 (1995).
\bibitem{ghx} C.S. Wood et al., Science {\bf 275}, 1759 (1997).
\bibitem{hw} W.C. Haxton and C.E. Wieman, Ann. Rev. Nucl. Part. Sci. {\bf 51}, 261 (2001).
\bibitem{fm} V.V. Flambaum and D.W. Murray, Phys. Rev. C {\bf 56}, 1641 (1997).
\bibitem {hlrm} W.C. Haxton, C.-P. Liu, and M.J. Ramsey-Musolf, Phys. Rev. Lett. {\bf 86} (2001) 5247.
\bibitem{ddh} B. Desplanques, J.F. Donoghue, and B.R. Holstein, Ann. Phys. (NY) {\bf 124}, 449 (1980).
\bibitem{btn} G. Barton, Nuovo Cim. {\bf 19}, 561 (1961).
\bibitem{pkl} B.R. Holstein, Phys. Rev. {\bf D23}, 1618 (1981).
\bibitem{dze} V.M. Dubovik and S.V. Zenkin, Ann. Phys. (NY) {\bf 172}, 100 (1986).
\bibitem{fcd} G.B. Feldman, G.A. Crawford, J. Dubach, and B.R. Holstein, Phys. Rev. {\bf C43}, 863 (1991).
\bibitem{meissner} U. G. Meissner and H. Weigel, Phys. Lett. {\bf B447}, 1 (1999).
\bibitem{lhk} H.-J. Lee, C. H. Hyun, and H.-C. Kim, arXiv:1203.4769 (to be published in Phys. Lett. B).
\bibitem{zhun} S.-L. Zhu, Phys. Rev. {\bf D79}, 116002 (2009).
\bibitem{wasem} J. Wasem, Phys. Rev. {\bf C85}, 022501(R) (2012).
\bibitem{bs} S.R. Beane and M.J. Savage, Nucl. Phys. {\bf B636}, 291 (2002).
\bibitem{tri} V.A. Vesna et al., Phys. Rev. {\bf C77}, 035501 (2008).
\bibitem{SNS} R.C. Gillis et al., J. of Phys.: Conf. Series {\bf 239}, 012012 (2010).
\bibitem{snow} M. Snow, private communication.
\bibitem{dm} D. E. Driscoll and G. A. Miller, Phys. Rev. {\bf C39}, 1951 (1989) and {\bf C40}, 2159 (1989).
\bibitem{carlson} J.A. Carlson, R. Schiavilla, V.R. Brown, and B.F. Gibson, Phys. Rev C {\bf 65}, 035502 (2002).
\bibitem{Lepage} P. Lepage, arXiv:nucl-th/9706029.
\bibitem{EFTStar} T. S. Park et al., Phys. Rev. C {\bf67}, 055206 (2003).
\bibitem{ramsey}W. D. Ramsey, Czech. J. Phys. {\bf 54} (2004) B207 (arXiv:nucl-ex/0401028).
\bibitem{CalLat} CaliforniaLattice Collaboration, J. Wasem {\it et al.}, private communication.
\bibitem{zhu} S.-L. Zhu et al., Nucl. Phys. {\bf A748}, 435 (2005).
\bibitem{dem} This is in the spirit of the approach suggested by B. Desplanques and J. Missimer, Nucl Phys. {\bf A300}, 286 (1978).
\bibitem{dan} G.S. Danilov, Phys. Lett. {\bf 18}, 40 (1965); Phys. Lett. {\bf B35}, 579 (1971); Sov. J. Nucl. Phys. {\bf 14}, 443 (1972).
\bibitem{gir} L. Girlanda, Phys. Rev. {\bf C77}, 067001 (2008).
\bibitem{phi} D.R. Phillips, M.R. Schindler, and R.P. Springer, Nucl. Phys. {\bf A822}, 1 (2009).
\bibitem{simonius} M. Simonius, Nucl. Phys. {\bf A220}, 269 (1974).
\bibitem{henley1975} E. M. Henley and F. R. Krejs, Phys. Rev. D {\bf11}, 605 (1975).
\bibitem{oka1981} T. Oka, Prog. Theor. Phys. {\bf 66}, 977 (1981).
\bibitem{flambaum} V. F. Dmitriev, V. V. Flambaum, O. P. Sushkov, and V. B. Telitsin, Phys. Lett. B {\bf 125}, 1 (1983).
\bibitem{dbg} B. Desplanques, J.J. Benayoun, and C. Gignoux, Nucl. Phys. {\bf A324}, 221 (1979);B. Desplanques and J. Benayoun, Nucl. Phys. {\bf A458}, 689 (1986).
\bibitem{avenier} M. Avenier, J. F. Cavaignac, D. Koang, B. Vignon, R. Hart, and R. Wilson, Phys. Lett. B. {\bf 137}, 125 (1984).
\bibitem{nsd} R. Schiavilla et al., Phys. Rev. {\bf C78}, 014002 (2008); Erratum Phys. Rev. {\bf C83}, 029902 (2011).
\bibitem{hg} H. W. Greisshammer, M. R. Schindler, and R. P. Springer, Eur. Phys. J. {\bf 48}, 7 (2012).

\end{thebibliography}
\end{document}